\documentclass[11pt,a4paper]{article}
\usepackage[T1]{fontenc}
\usepackage[utf8]{inputenc}
\usepackage{lmodern}
\usepackage[margin=27mm]{geometry}
\usepackage{amsmath,amssymb,amsthm,mathtools}
\usepackage{microtype,xcolor}
\usepackage[colorlinks=true,linkcolor=blue!55!black,citecolor=blue!55!black,urlcolor=blue!55!black]{hyperref}
\hypersetup{pdftitle={Singularities of the cosmological time and Cauchy Busemann functions: invariance, uniqueness and global timelike propagation}}
\newtheorem{theorem}{Theorem}[section]
\newtheorem{lemma}[theorem]{Lemma}
\newtheorem{proposition}[theorem]{Proposition}
\newtheorem{corollary}[theorem]{Corollary}
\theoremstyle{definition}
\newtheorem{definition}[theorem]{Definition}
\theoremstyle{remark}
\newtheorem{remark}[theorem]{Remark}
\DeclareMathOperator{\dom}{dom}
\DeclareMathOperator{\co}{co}
\DeclareMathOperator{\Sing}{Sing}
\DeclareMathOperator*{\argmax}{arg\,max}
\newcommand{\R}{\mathbb R}
\newcommand{\dd}{\mathrm d}
\newcommand{\cS}{\mathcal S}
\newcommand{\cF}{\mathcal F}
\newcommand{\cR}{\mathcal R}
\newcommand{\cE}{\mathcal E}
\newcommand{\normL}[1]{\lvert #1\rvert_L}
\newcommand{\norma}[1]{\lvert #1\rvert_a}

\title{\Large Singularities of the cosmological time and Cauchy Busemann functions:\\ invariance, uniqueness and global timelike propagation}
\author{E. Minguzzi\footnote{Dipartimento di Matematica, Universit\`a degli
Studi di Pisa, Via B. Pontecorvo 5, I-56127 Pisa, Italy. E-mail:
ettore.minguzzi@unipi.it, ORCID:0000-0002-8293-3802}}
\date{}

\begin{document}
\maketitle
\begin{abstract}
\noindent Let $b$ be the future Cauchy Busemann function of a compact Cauchy hypersurface in a globally hyperbolic, future timelike complete spacetime, and let $\cS$ be its singular set. The convex hulls of the initial tangents of the calibrated rays generate a closed cone structure $C^b$. We prove that $\cS$ is past invariant for $C^b$, with an exponential lower bound on the strength of the singularities, and that past-directed $C^b$-causal curves are unique: they are the trajectories of a jointly locally Lipschitz past semi-flow, which preserves $\cS$ and may merge trajectories. Hence every singular point is the future endpoint of a unique past-inextendible, locally uniformly timelike singular curve. The points whose trajectory never enters $\cS$ are those of the calibrated timelike lines, which exist whenever the cosmological time is regular; every other point is either an initial point, at which $b$ admits no smooth lower support, or an interior point of exactly one maximal calibrated ray. The initial points form a null set on every $C^2$ spacelike hypersurface, and the differential of $b$ is locally Lipschitz on each set where the calibrated past extension has a fixed positive temporal length. In a counterexample to Bartnik's splitting conjecture every trajectory enters $\cS$, uniformly on compact sets, and the domain of $b$ contains no Cauchy hypersurface. The fibres of the semi-flow are single points exactly over the regular points, and they are cellular under a compactness condition. Explicit models show the sharpness of several statements, including the existence of spacelike intrinsic Lax--Oleinik maximizing arcs. No curvature condition and no past completeness are assumed. By time duality, the singularities of a regular cosmological time, once formed, persist toward the future along unique timelike curves.
\end{abstract}

\newpage
\tableofcontents
\newpage

\section{Introduction}\label{sec:intro}

\subsection{Background and motivation}

The galaxies are not distributed uniformly. On the scale of tens of
megaparsecs they trace a network of walls, filaments and compact nodes
which surround nearly empty voids, the cosmic web \cite{bond96}. A simple
and successful description of the formation of this network is the
adhesion model of Gurbatov, Saichev and Shandarin
\cite{gurbatov89,shandarin89}, which extends the Zel'dovich approximation
\cite{zeldovich70} beyond the first crossing of the trajectories of the
matter. In this model the matter is a pressureless fluid whose velocity
potential solves, in suitable comoving variables, the Hamilton--Jacobi
equation of free motion in the limit of vanishing viscosity, namely
Burgers' equation. The solution is given by an explicit variational
formula, and it fails to be differentiable on a set of shocks, where
streams of matter with different velocities meet and stick together. The
shocks are the walls of the web; they intersect along the filaments, which
meet at the nodes \cite{vergassola94}. Three features of this picture are
essential. The shocks, once
formed, do not disappear. They move with a velocity which is determined by
the streams which meet there, the mean of their velocities in the simplest
case. Finally, the matter which has entered the shocks does not leave them:
it flows along the walls into the filaments and along the filaments into
the nodes, so that trajectories merge but do not branch.

The adhesion model is Newtonian, and it rests on an approximation of the
gravitational interaction. The purpose of this paper is to show that its
three features are exact properties of the singularities, namely of the
sets of non-differentiability points, of two functions which are
canonically associated with a spacetime, in general relativity and beyond:
no field equation, no symmetry and no curvature condition is assumed.
Throughout the paper the singularities are those of these two functions,
not those of the spacetime.

The first function is the cosmological time $\tau$ of Andersson, Galloway
and Howard \cite{andersson98}, which assigns to every event the supremum
of the proper times of the observers which reach it from the past. When it
is regular, every event is reached by a maximal timelike geodesic which
emanates from the initial singularity, and $\tau$ is a locally semiconvex
function which satisfies $g(\nabla\tau,\nabla\tau)=-1$ at its
differentiability points, the relativistic Hamilton--Jacobi equation of
free fall. In the Friedmann--Lema\^itre--Robertson--Walker models these
geodesics are the worldlines of the comoving matter. In general they are
the worldlines of the observers which have aged maximally since the
initial singularity, and where $\tau$ is differentiable they form an
irrotational geodesic congruence, the counterpart of the streams of the
adhesion model, with the gravitational field fully taken into account. The singular set $\cS_\tau$
of $\tau$ consists of the events which are reached by more than one of
them, and it is the counterpart of the shocks.

The second function is the future Cauchy Busemann function $b=b^+$ of a
compact Cauchy hypersurface $\Sigma$, introduced in \cite{minguzzi26e} for
future timelike complete spacetimes. Its finiteness domain, the future
cosmological development, and its boundary, the future cosmological
horizon, do not depend on $\Sigma$. The function $b$ is locally Lipschitz
and locally semiconcave, its graph is a Cauchy horizon in the product
spacetime $M\times\R$, and through every point of its domain passes a
calibrated ray: a future complete timelike geodesic along which $b$
increases by the Lorentzian length, namely an observer who ages maximally
into the far future. The function $b$ is differentiable exactly at the
points from which a single calibrated ray starts, and its singular set
$\cS$ consists of the events from which several calibrated rays start. The
two functions are related by time duality: in the spacetime with the
opposite time orientation, $-\tau$ has the structural properties of $b$
(Section~\ref{sec:cosmo}). The maximal geodesics from the initial
singularity converge on $\cS_\tau$, while the calibrated rays diverge from
$\cS$.

For semiconcave solutions of Hamilton--Jacobi equations, singularities
propagate along generalized characteristics, and the propagation of
singularities and the structure of the cut locus have been studied in
detail by Cannarsa, Cheng and coauthors \cite{CC,CCHW}. For the Cauchy
Busemann function a local version was obtained in
\cite[Theorem~5.21(d)]{minguzzi26e}: every singular point starts a Lipschitz
singular curve. The global propagation of singularities toward the past,
along timelike curves, was conjectured there. In this paper we prove it,
and we develop a global theory of the singular set. Its central object is
the calibrated cone structure: at each point $x$ of the domain of $b$, the
convex cone $C^b_x$ generated by the initial unit tangents of the calibrated
rays starting at $x$. This cone is a single ray at differentiability points
and a fan at singular points, and it is a closed cone structure in the
sense of \cite{minguzzi17}. We study its causal curves, oriented toward the
past. The Lorentzian setting differs from the Hamilton--Jacobi theory in a
notable respect: intrinsic Lax--Oleinik maximizing arcs can be spacelike
(Appendix~\ref{app:example}), whereas the causal curves of $C^b$ are always
timelike.

We state and prove the results for $b$, toward the past, and we transfer
them to $\tau$, toward the future, by time duality
(Theorem~\ref{thm:cosmo}). For the cosmological time they give the three
features recalled above. The singular set $\cS_\tau$ is invariant toward
the future, with an exponential lower bound on the strength of the
singularities. Every singular point moves toward the future along a unique
timelike curve, whose velocity is proportional to the vector of maximal
Lorentzian norm in the convex hull of the arriving directions, which is
their mean when they are two (Remark~\ref{rem:midpoint}). Finally, the
maximal geodesics from the initial singularity are the regular part of the
trajectories of a locally Lipschitz future semi-flow, which preserves
$\cS_\tau$ and merges trajectories without branching. In the adhesion model
the analogous facts follow from the explicit formula for the solution;
here no such formula is available, and the proofs rest on the
semiconcavity of $b$, on its support functions and on the causality theory
of the cone structure $C^b$.

The Cauchy Busemann function was introduced in \cite{minguzzi26e} in
connection with Bartnik's splitting conjecture, and the consequences of
our results for this conjecture are collected in
Section~\ref{sec:bartnik}.

\subsection{Setting and notation}\label{sec:setting}

Let $(M,g)$ be a smooth, time-oriented, globally hyperbolic spacetime of
dimension $n+1$ with compact Cauchy hypersurfaces. Assume future timelike
geodesic completeness. Fix a compact Cauchy hypersurface $\Sigma$ and let
$b$ be its future Cauchy Busemann function, or any function differing from
it by an additive constant. Put
\[
 O=\dom b,\qquad \cS=\Sing(b)\subset O.
\]
This convention includes the canonical function
$b^{\textnormal{\textsc{m}}}$ for spacetimes with regular cosmological time:
by \cite[Proposition~5.23(b)]{minguzzi26e},
$b^{\textnormal{\textsc{m}}}=b^{T_a}+a$ for every $a>0$, where $T_a$ is the
cosmological-time level of value $a$ and $b^{T_a}$ is its Cauchy Busemann
function. All statements below are invariant under addition of a constant
to $b$. The analytic estimates are local in $O$; the global continuation
arguments also use global hyperbolicity and $J^-(O)\subset O$.

Fix a smooth Cauchy temporal function $T:M\to\R$, increasing toward the
future, and a complete auxiliary Riemannian metric, with norm $\norma{\cdot}$
and distance $d_a$. For a timelike vector write
$\normL{v}=\sqrt{-g(v,v)}$.

For $x\in O$, let $U_x$ be the set of initial unit tangents of future
$b$-calibrated rays. For a subset $A$ of a vector space, $\co A$ denotes its
convex hull; thus $\co U_x\subset T_xM$ is the set of finite convex
combinations of elements of $U_x$. We use the following facts from
\cite{minguzzi26e}:
\begin{enumerate}
\item $O$ is open and $J^-(O)\subset O$ (Proposition~3.7).
\item $b$ is locally Lipschitz and locally semiconcave on $O$; the sets
      $U_x$ are nonempty, and their union over a compact base set is compact
      in $TM$ (Theorem~5.12).
\item For $u\in U_x$ and $\sigma>0$,
\begin{equation}\label{eq:support}
 b(y)\le b(x)+\sigma-d(y,\exp_x(\sigma u))
 \quad\text{if }y\ll\exp_x(\sigma u).
\end{equation}
The right-hand side is a smooth upper support near $x$, with differential
$-g_x(u,\cdot)$ there (Theorem~5.12(iii)).
\item At differentiability points, $-\nabla b(x)\in U_x$ and
      $g(\nabla b,\nabla b)=-1$ (Lemma~5.16).
\item The rushing inequality is
\begin{equation}\label{eq:rushing}
 b(y)-b(x)\ge d(x,y)\qquad(x\le y,\ x,y\in O).
\end{equation}
\end{enumerate}

The superdifferential $D^+b(x)\subset T^*_xM$ is the set of covectors $P$
such that, in some chart around $x$, and then in every chart,
\[
 b(y)\le b(x)+P(y-x)+o(|y-x|)\qquad(y\to x).
\]
A reachable differential at $x$ is a limit of differentials $\dd b_{x_j}$ at
differentiability points $x_j\to x$. Since $b$ is locally semiconcave,
$D^+b(x)$ is a nonempty compact convex set, $x\mapsto D^+b(x)$ is upper
semicontinuous, $D^+b(x)$ is the convex hull of the reachable
differentials, and $b$ is differentiable at $x$ if $D^+b(x)$ is a singleton
\cite[Proposition~2.1(a)--(d)]{CC}; conversely $D^+b(x)=\{\dd b_x\}$ at
differentiability points. Moreover, in a chart and locally uniformly,
\begin{equation}\label{eq:semiconc}
 b(y)\le b(x)+P(y-x)+A|y-x|^2\qquad(P\in D^+b(x))
\end{equation}
\cite[Proposition~2.2]{CC}. No propagation theorem from \cite{CC} is used.

We put
\begin{equation}\label{eq:strength}
 \vartheta(x)=\max_{v\in\co U_x}\normL{v},\qquad
 C^b_x=\R_{>0}\co U_x,\qquad
 \cF_x=\left\{\frac{v}{\dd T_x(v)}:v\in\co U_x\right\}.
\end{equation}
A past-directed $C^b$-causal curve is a locally absolutely continuous curve
$x$ in $O$ with $x'\in-C^b_x$ almost everywhere. Since $\co U_x$ consists of
future timelike vectors of Lorentzian norm at least one, and these sets
form a compact family over compact subsets of $O$ by fact~2, the velocity
of such a curve is nonzero almost everywhere; reparametrized by $T$, the
curve is locally Lipschitz and locally uniformly timelike, and proper time
is a locally bi-Lipschitz parameter. Here locally uniformly timelike means
that on each compact parameter interval the almost-everywhere velocities
belong to a compact subset of the timelike cone bundle.

\subsection{Main results}

\begin{theorem}[Past invariance with exponential persistence]\label{thm:invariance}
Let $x:[0,L]\to O$ be a past-directed $C^b$-causal curve, parametrized by
proper time. There is a constant $E\ge0$, depending only on a compact
neighborhood of $x([0,L])$, such that
\begin{equation}\label{eq:main}
 \vartheta(x(t))-1\ge
 e^{-E(t-s)}\bigl(\vartheta(x(s))-1\bigr)
 \qquad(0\le s\le t\le L).
\end{equation}
In particular, if $x(s)\in\cS$, then $x(t)\in\cS$ for every $t\ge s$, so
that $\cS$ is a past set for the causal relation of $C^b$.
\end{theorem}

\begin{theorem}[Global timelike singular propagation]\label{thm:main}
For every $p\in\cS$ there is a locally Lipschitz curve
\[
 \gamma:(-\infty,T(p)]\longrightarrow\cS,
 \qquad \gamma(T(p))=p,\qquad T(\gamma(\tau))=\tau,
\]
which is locally uniformly future timelike and past inextendible, and
satisfies
\begin{equation}\label{eq:main-inclusion}
 \dot\gamma(\tau)\in\cF_{\gamma(\tau)}
 \quad\text{for almost every }\tau,
\end{equation}
and, for every $a<T(p)$,
\[
 \inf_{a\le\tau\le T(p)}\vartheta(\gamma(\tau))>1.
\]
Every past-inextendible past-directed $C^b$-causal curve starting at $p$,
parametrized by $T$, has these properties.
\end{theorem}

\begin{theorem}[Past uniqueness and the past semi-flow]\label{thm:unique}
Under the standing assumptions:
\begin{enumerate}
\item Two past-directed $C^b$-causal curves with the same initial point
      coincide on their common parameter domain when both are parametrized
      by proper time, or both by the drop of $T$. Equivalently, their
      oriented images agree up to increasing reparametrization until one
      of them ends.
\item There is a unique map $\Psi:[0,\infty)\times O\to O$ such that, for
      every $p\in O$, $s\mapsto\Psi_s(p)$ is a past-directed $C^b$-causal
      curve with
      \[
       \Psi_0(p)=p,\qquad T(\Psi_s(p))=T(p)-s .
      \]
      The map $\Psi$ is jointly locally Lipschitz,
      $\Psi_{s+t}=\Psi_s\circ\Psi_t$ for $s,t\ge0$, and
      $\Psi_s(\cS)\subset\cS$ for $s\ge0$. For every
      $p\in O$ and every $s\ge0$ the right derivative
      \begin{equation}\label{eq:generator}
       \frac{\dd^+}{\dd s}\Psi_s(p)
       =-\frac{V(\Psi_s(p))}{\dd T\bigl(V(\Psi_s(p))\bigr)}
      \end{equation}
      exists, and it is right-continuous in $s$. The oriented trajectories
      of $\Psi$ do not depend on the choice of $T$.
\end{enumerate}
\end{theorem}

In particular, the curve of Theorem~\ref{thm:main} is unique, namely
$\gamma(\tau)=\Psi_{T(p)-\tau}(p)$. No smoothness of it is asserted, and
past inextendibility does not assert infinite past proper time. Past
uniqueness does not imply injectivity of $\Psi_s$, and uniqueness fails in
the future direction at singular points, from which several calibrated
rays start (Remark~\ref{rem:merge}). We write $\Psi$ because $\Phi$ denotes
the canonical calibrating semi-flow in \cite[Section~5.1]{minguzzi26e}, which
moves along the calibrated rays toward the future.

\subsection{Trajectories, initial points and fibres}

The remaining results describe the trajectories of $\Psi$ and its fibres.
Let $s_0(p)\in[0,\infty]$ be the drop of $T$ after which the trajectory of
$p$ enters $\cS$.
\begin{itemize}
\item {\em Regular phase.} Before $s_0(p)$ the trajectory runs backward
      along a calibrated ray through $p$; afterwards it remains in $\cS$
      (Theorem~\ref{thm:regular}). The function $s_0$ is upper
      semicontinuous, but in general not lower semicontinuous
      (Proposition~\ref{prop:entry} and Remark~\ref{rem:Qdisc}).
\item {\em Calibrated lines and initial points.} The points with
      $s_0=\infty$ are exactly those of the $b$-calibrated timelike lines;
      they form a closed subset $\mathcal L_b$ of the domain of $b$. The
      points with $s_0=0$ are the initial points, namely the points which
      are not interior points of a calibrated ray; they are also the points
      at which $b$ admits no smooth lower support, and they form a set
      $\cE$ with $\cS\subset\cE\subset\overline\cS$, of measure zero on
      every $C^2$ spacelike hypersurface. Every other point is an interior
      point of exactly one maximal calibrated ray, which starts on $\cE$
      (Corollary~\ref{cor:structure}, Theorems~\ref{thm:support}
      and~\ref{thm:measure}).
      Calibrated lines exist as soon as $b+d(\cdot,\Lambda)$
      stays bounded along a sequence escaping to the past, for a compact
      Cauchy hypersurface $\Lambda$; in particular they exist whenever the
      cosmological time is regular (Proposition~\ref{prop:pastdefect} and
      Corollary~\ref{cor:depth}).
\item {\em Capture.} Away from $\mathcal L_b$, trajectories are captured
      by $\cS$ after a drop of $T$ which is bounded on compact sets
      (Proposition~\ref{prop:capture}).
\item {\em Regularity.} On each set $\{s_0\ge\delta\}$, $\delta>0$, the
      differential of $b$ is locally Lipschitz, with constant of order
      $1/\delta$, and the forward calibrated maps are locally bi-Lipschitz;
      in particular $\mathcal L_b$ carries a locally Lipschitz flow along
      the calibrated lines. Quantitatively, $s_0(p)$ is comparable to the
      reciprocal of the constant of the best quadratic lower support of $b$
      on the time slice through $p$ (Theorems~\ref{thm:gradient}
      and~\ref{thm:quant}, Corollary~\ref{cor:forward}).
\item {\em Fibres.} The fibre of $\Psi_s$ over $y$ is the slice of the
      future reachable set of $y$ for $C^b$. It is a single point if $y$ is
      regular, it has at least two points if $y\in\cS$ and $s>0$, and it is
      a cellular set when the compact future region
      $J^+(y)\cap T^{-1}([T(y),T(y)+s])$ lies in the domain of $b$. In the
      latter case it lies in the causal future of $y$ for a metric with
      cones strictly narrower than the light cones: the fibre dilates at a
      speed lower than the speed of light (Remark~\ref{rem:speed}). If a
      level $T^{-1}(c)$ lies in the domain of $b$, the restriction of each
      $\Psi_s$ to it is a uniform limit of diffeomorphisms
      (Theorem~\ref{thm:fibres}).
\end{itemize}

\subsection{A wide cone with a single past curve}

Theorem~\ref{thm:invariance}
concerns every past-directed $C^b$-causal curve, while
Theorem~\ref{thm:unique} shows that from each point there is only one. This
is compatible with the width of the cones. At a singular point $x$ the cone
$C^b_x$ is a fan, spanned by the initial directions of several calibrated
rays, whereas at a regular point it is the single ray spanned by the initial
direction of the unique calibrated ray. A $C^b$-causal curve must satisfy the
cone condition at almost every point it traverses, not only at its initial
point, and only one direction of the fan can be sustained in this way.
Indeed, along such a curve $b$ must decrease at the same rate when compared
with the support function of each of the competing calibrated rays, and this
forces the velocity to be a negative multiple of the balanced vector $V$
(Lemma~\ref{lem:rigidity}). In the model of Remark~\ref{rem:merge},
$b=At-a|z|$, the cone at a point of the singular axis $z=0$ is the fan
between the past directions of the two calibrated rays, but a past-directed
curve which leaves the axis toward either side enters a region where the
only past direction points back to the axis; the only consistent motion is
along the axis, in the direction $-V$. Uniqueness rests on a convexity
property: for the reversed time orientation, suitably parametrized
past-directed $C^b$-causal curves are gradient curves of the function
$e^{-kb}$, which is locally convex along causal geodesics
(Section~\ref{sec:unique}). Gradient curves of convex functions are unique
even where the function has a corner and its subdifferential is
multivalued: for $f(x)=|x|$ on the real line the subdifferential at $0$ is
$[-1,1]$, yet every solution of $\dot x\in-\partial f(x)$ with $x(0)=0$
satisfies $(x^2)'=-2|x|\le0$, hence stays at $0$. In the Lorentzian
setting the element of maximal, rather than minimal, norm is selected,
because the reverse Cauchy--Schwarz inequality reverses the inequalities.
In the future direction the width of the cones does produce several curves:
from a singular point at least two calibrated rays start, and the set of
later points which $\Psi$ maps to a singular point has more than one point
(Theorem~\ref{thm:fibres}); in the model it is a segment, since
future-directed $C^b$-causal curves may run along the axis before leaving
it. Past-directed curves are thus funnelled into $\cS$ and merge there, while
future-directed ones branch at the points of $\cS$.

\subsection{Time duality: the cosmological time}

The same results hold, with the time orientation reversed, for the
cosmological time
$\tau(q)=\sup\{\ell(\gamma):\gamma$ past-directed causal curve starting at
$q\}$ of Andersson, Galloway and Howard \cite{andersson98}, when it is
regular, namely $\tau<\infty$ on $M$ and $\tau\to0$ along every
past-inextendible causal curve. Here no completeness and no compactness of
the Cauchy hypersurfaces is assumed. A {\em rest curve} of $q$ is a future
directed unit speed timelike geodesic $\gamma:(0,\tau(q)]\to M$ with
$\gamma(\tau(q))=q$ and $\tau(\gamma(t))=t$, a maximal geodesic from the
initial singularity; every point has at least one \cite{andersson98}. Let
$R_q$ be the set of the final unit tangents of the rest curves of $q$, let
$\cS_\tau$ be the set of non-differentiability points of $\tau$, and define
$\vartheta_\tau$, $V_\tau$, $C^\tau=\R_{>0}\co R$ and $\cF^\tau$ as in
\eqref{eq:strength}, with $U$ replaced by $R$. A future-directed
$C^\tau$-causal curve is a locally absolutely continuous curve $x$ with
$x'\in C^\tau_x$ almost everywhere.

\begin{theorem}[Future propagation for the cosmological time]\label{thm:cosmo}
Let $(M,g)$ have regular cosmological time $\tau$ and let $T$ be a smooth
Cauchy temporal function.
\begin{enumerate}
\item Along every future-directed $C^\tau$-causal curve $x:[0,L]\to M$
      parametrized by proper time,
      $\vartheta_\tau(x(t))-1\ge e^{-E(t-s)}(\vartheta_\tau(x(s))-1)$ for
      $0\le s\le t\le L$, with $E$ depending only on a compact
      neighbourhood of $x([0,L])$. In particular $\cS_\tau$ is a future set
      for the causal relation of $C^\tau$.
\item There is a unique map $\Psi^\tau:[0,\infty)\times M\to M$ such that
      $s\mapsto\Psi^\tau_s(p)$ is a future-directed $C^\tau$-causal curve
      with $\Psi^\tau_0(p)=p$ and $T(\Psi^\tau_s(p))=T(p)+s$. It is jointly
      locally Lipschitz, $\Psi^\tau_{s+t}=\Psi^\tau_s\circ\Psi^\tau_t$,
      $\Psi^\tau_s(\cS_\tau)\subset\cS_\tau$, its right derivative is
      $V_\tau/\dd T(V_\tau)$, and its oriented trajectories do not depend
      on $T$. Two future-directed $C^\tau$-causal curves with the same
      initial point have the same oriented image until one of them ends.
\item Every $p\in\cS_\tau$ is the past endpoint of a unique
      future-inextendible $C^\tau$-causal curve; it is contained in
      $\cS_\tau$, locally uniformly timelike, and
      $\inf\vartheta_\tau>1$ on each of its compact subarcs. For every
      Cauchy hypersurface $\Sigma$ with $p\in J^-(\Sigma)$,
      $\cS_\tau\cap\Sigma\cap J^+(p)\ne\varnothing$, and
      $\mathcal H^n(\cS_\tau\cap A)=0$ for every achronal set $A$. If
      $\cS_\tau\ne\varnothing$, every connected component of $\cS_\tau$
      is unbounded above for $T$.
\item If $\tau$ is differentiable at every point of a Cauchy hypersurface
      $\Sigma$, it is differentiable on $J^-(\Sigma)$ and
      $C^{1,1}_{\mathrm{loc}}$ on $I^-(\Sigma)$.
\end{enumerate}
\end{theorem}

The singular points of $\tau$ are the events reached by more than one
maximal geodesic from the initial singularity. Theorem~\ref{thm:cosmo}
says that, once formed, they never disappear: each of them moves toward
the future along a unique timelike curve, with velocity proportional to the
maximal-norm vector of the convex hull of the arriving directions
(Remark~\ref{rem:midpoint}). In contrast with $\cS$, whose components are
unbounded toward the past, $\cS_\tau$ may be empty before some time and
appear afterwards. The proof, in Section~\ref{sec:cosmo}, checks that
$-\tau$ satisfies the standing facts in the time-reversed spacetime.

\subsection{Consequences for Bartnik's conjecture}\label{sec:bartnik}

Bartnik's splitting conjecture \cite{bartnik88} asserts that a globally
hyperbolic spacetime with compact Cauchy hypersurfaces, which is timelike
geodesically complete and satisfies the timelike convergence condition
(TCC), splits isometrically as a static product $(\R\times S,-\dd t^2+h)$.
By the Lorentzian splitting theorem \cite{eschenburg88,galloway89,newman90}
it suffices to find a complete timelike line, so that the conjecture is a
question on the existence of maximizing timelike geodesics. If $b$ is
$C^1$ on its domain, a timelike complete spacetime contains a timelike
line \cite[Theorem~5.17]{minguzzi26e}, so in a counterexample to the
conjecture the singular set $\cS$ of $b$ is nonempty
\cite[Remark~5.20]{minguzzi26e}, and its structure is relevant to the
conjecture.

Let $(M,g)$ be a counterexample to Bartnik's conjecture: a timelike
complete, globally hyperbolic spacetime with compact Cauchy hypersurfaces,
satisfying TCC, which does not split. By the Lorentzian splitting theorem
it contains no timelike line, and our results give the following picture
of its future cosmological development $O$.
\begin{enumerate}
\item The singular set $\cS$ is nonempty, and every singular point is the
      future endpoint of a past-inextendible timelike singular curve, which
      meets every Cauchy hypersurface in its past
      (Theorem~\ref{thm:main} and Corollary~\ref{cor:geometry}).
\item Every trajectory of the past semi-flow enters $\cS$ after a finite drop
      of $T$, uniformly for initial points in compact subsets of $O$
      (Corollary~\ref{cor:nolines}).
\item Every point of $O$ is either an initial point or an interior point of
      exactly one maximal calibrated ray, which starts at an initial point.
      Thus all the maximally aging observers are born on the closure of
      $\cS$ (Corollary~\ref{cor:nolines}).
\item The domain $O$ contains no Cauchy hypersurface
      (Corollary~\ref{cor:cauchyline}).
\item For every compact Cauchy hypersurface $\Lambda$, the
      function $b+d(\cdot,\Lambda)$ diverges at past infinity: its sublevel
      sets in $J^-(\Lambda)$ are compact
      (\cite[Corollary~7.8]{minguzzi26e} and Corollary~\ref{cor:depth}).
\end{enumerate}
None of these statements uses the curvature condition, which enters only
through the splitting theorem.

\subsection{Relation to other work}

The past invariance of $\cS$ and the global propagation theorem answer the
question raised after \cite[Theorem~5.21]{minguzzi26e}, and the selection of
the maximal-norm vector made by the cone condition gives the inclusion
proposed there (Remark~\ref{rem:inclusion}). Existence of causal curves
and limit curve arguments come from the theory of closed cone structures
\cite{minguzzi17}. From the theory of semiconcave functions we use only
basic facts on superdifferentials \cite{CC}; in particular, no Tonelli or
Lax--Oleinik representation of $b$ is used, and Appendix~\ref{app:example}
shows that such representations behave differently in the Lorentzian
setting. Our characterization of the initial points identifies them with
the analytic cut locus of \cite{CCHW}. Under the identification of $b$ with
a horizon in $M\times\R$, the initial points correspond to the past
endpoints of the null generators, a set studied for general horizons by
Chru\'sciel, Fu, Galloway and Howard \cite{CFGH}. The uniqueness proof
compares chronologically ordered gradient curves of a causally convex
function, in the spirit of the Lorentzian gradient flows considered in
\cite[Section~2.3]{Almost}. The connectedness of the fibres can also be
derived from a Kneser-type theorem for differential inclusions \cite{KK},
while their cellularity is in the sense of \cite{Uspenskij}.

\subsection{Organization}

Section~\ref{sec:strength} introduces the strength, the maximal-norm vector
and the calibrated cone structure, and proves the velocity rigidity of
cone-causal curves. Section~\ref{sec:propagation} proves
Theorems~\ref{thm:invariance} and~\ref{thm:main} and collects consequences
for the geometry of $\cS$. Section~\ref{sec:unique} proves
Theorem~\ref{thm:unique}, including the Lipschitz regularity of the
semi-flow, and Section~\ref{sec:spheres} gives a constructive proof of
existence by maximization on past spheres. Section~\ref{sec:rays} studies
calibrated rays, initial points and calibrated lines, and
Section~\ref{sec:fibres} the fibres of the semi-flow and the forward
calibrated maps. Section~\ref{sec:cosmo} proves Theorem~\ref{thm:cosmo} by
time duality, and Section~\ref{sec:conclusions} contains conclusions and
open problems. Appendix~\ref{sec:flat} treats flat affine models, which show
the sharpness of some of the statements, and Appendix~\ref{app:example}
contains an exact spacelike intrinsic arc.

\section{The calibrated cone structure}\label{sec:strength}

\subsection{Strength and the maximal-norm vector}

\begin{lemma}\label{lem:strength}
For every $x\in O$,
\begin{equation}\label{eq:superdiff}
 D^+b(x)=\{-g_x(v,\cdot):v\in\co U_x\}.
\end{equation}
The function $\vartheta$ is upper semicontinuous, $\vartheta\ge1$, and
$\vartheta(x)=1$ exactly at the differentiability points of $b$.
Consequently $\{\vartheta\ge1+\kappa\}$ is closed in $O$ and contained in
$\cS$ for every $\kappa>0$.
\end{lemma}

\begin{proof}
Each covector $-g_x(u,\cdot)$, $u\in U_x$, belongs to $D^+b(x)$ by
\eqref{eq:support}; convexity gives one inclusion in
\eqref{eq:superdiff}. Conversely, $D^+b(x)$ is the convex hull of the
reachable differentials \cite[Proposition~2.1(d)]{CC}. A reachable
differential is a limit of $\dd b_{x_j}=-g_{x_j}(u_j,\cdot)$ with
$u_j=-\nabla b(x_j)\in U_{x_j}$ \cite[Lemma~5.16]{minguzzi26e} and $x_j\to x$.
Compactness of calibrated tangents gives a subsequence converging to
$u\in U_x$, proving the opposite inclusion.

Every convex combination of future unit timelike vectors has Lorentzian
norm at least one. If two distinct unit vectors occur, their midpoint has
norm strictly greater than one. Thus $\vartheta=1$ exactly when $U_x$,
equivalently $D^+b(x)$, is a singleton, which is equivalent to
differentiability \cite[Proposition~2.1(c)]{CC}.

If $x_j\to x$, choose maximizing $v_j\in\co U_{x_j}$. By
Carath\'eodory's theorem each $v_j$ is a convex combination of at most
$n+2$ elements of $U_{x_j}$. Passing to subsequences in the compact tangent
family and in the coefficient simplex gives a limit $v\in\co U_x$.
Along a subsequence realizing the upper limit,
$\limsup_j\vartheta(x_j)=\normL{v}\le\vartheta(x)$.
\end{proof}

\begin{lemma}\label{lem:V}
For every $x\in O$ there is a unique $V(x)\in\co U_x$ with
$\normL{V(x)}=\vartheta(x)$.
\end{lemma}

\begin{proof}
If $v_1\ne v_2$ are maximizers, the reverse triangle inequality gives
$\normL{(v_1+v_2)/2}\ge\vartheta(x)$, with strict inequality unless $v_1$
and $v_2$ are positively collinear; equal norms exclude the latter case.
\end{proof}

\begin{lemma}\label{lem:F}
The sets $\cF_x$ are nonempty, compact and convex. In local
trivializations, $x\mapsto\cF_x$ is upper semicontinuous. Over compact
subsets of $O$, their union is compact and consists entirely of future
timelike vectors satisfying $\dd T(v)=1$.
\end{lemma}

\begin{proof}
On a compact base set, $\co U_x$ forms a compact family of future timelike
vectors, all of norm at least one. The continuous positive quantity
$\dd T(v)$ is bounded away from zero there. This proves compactness and
the closed-graph property after normalization, hence upper
semicontinuity. Convexity follows from
\[
 \cF_x=C^b_x\cap\{w:\dd T_x(w)=1\},
\]
as the positive cone over a convex set is convex. Convex combinations of
future timelike vectors are timelike, and compactness supplies a uniform
cone margin.
\end{proof}

\begin{remark}[Closed cone structures]\label{rem:cones}
By Lemma~\ref{lem:F}, $x\mapsto C^b_x$ is a closed cone structure on $O$ in
the sense of \cite[Definitions~2.2 and 2.3, Proposition~2.3]{minguzzi17}:
its fibres are closed, sharp, convex and nonempty cones, and it is upper
semicontinuous. It is narrower than the cone structure of $g$, $C^b<C^g$,
since $\co U_x$ consists of timelike vectors, and it is not proper: at
differentiability points $C^b_x$ is the ray spanned by $-\nabla b(x)$, the
initial direction of the unique calibrated ray. Past-directed
$C^b$-causal curves are the continuous causal curves of the reversed
structure, namely the solutions of the differential inclusion
\cite[Eq.~(2.2)]{minguzzi17}, and $\cF_x$ is the slice of $C^b_x$ on
$\{\dd T=1\}$.

Under the homeomorphism $x\mapsto(x,b(x))$ onto the Cauchy horizon $H$ of
the product spacetime $(M\times\R,g+\dd z^2)$ used in \cite{minguzzi26e},
$\R_{>0}\{(v,1):v\in\co U_x\}$ is the convex cone generated by the
semitangents of $H$, and its projection onto $T_xM$ is $C^b_x$. The
singular set $\cS$ corresponds to the non-differentiability set of $H$, so
Theorems~\ref{thm:invariance}--\ref{thm:unique} can be read on $H$ through
the projected cone structure. They are not statements on causal curves of
the product metric: by Lemma~\ref{lem:rigidity} below, the graph lift
$\widetilde x=(x,b(x))$ of a proper-time past-directed $C^b$-causal curve
has, almost everywhere,
\[
 \widetilde x'=\Bigl(-\frac{V}{\vartheta},-\vartheta\Bigr),\qquad
 (g+\dd z^2)(\widetilde x',\widetilde x')=\vartheta^2-1,
\]
so it is spacelike at almost every singular time, whereas every vector
$\lambda(v,1)$ of the semitangent cone has squared norm
$\lambda^2(1-\normL v^2)\le0$.
\end{remark}

\subsection{Velocity rigidity along cone-causal curves}\label{sec:rigidity}

\begin{lemma}[Velocity rigidity and exact defect]\label{lem:rigidity}
Along any past-directed $C^b$-causal curve $x$,
\begin{equation}\label{eq:rigidity}
 x'(t)=-\lambda(t)V(x(t)),\qquad \lambda(t)>0,
\end{equation}
at almost every parameter value. Moreover,
\begin{equation}\label{eq:exact}
 -(b\circ x)'(t)=\vartheta(x(t))\normL{x'(t)}
 \quad\text{almost everywhere}.
\end{equation}
In proper time these identities become
\begin{equation}\label{eq:proper}
 x'=-\frac{V(x)}{\vartheta(x)},\qquad
 -(b\circ x)'=\vartheta(x).
\end{equation}
\end{lemma}

\begin{proof}
Work at a parameter where $x$ and $b\circ x$ are differentiable and the
cone inclusion holds. Inequality \eqref{eq:semiconc}, applied with
increments of both signs, gives
\[
 (b\circ x)'=P(x')\qquad\text{for every }P\in D^+b(x).
\]
Write $x'=-\lambda v$ with $\lambda>0$ and $v\in\co U_x$. By
\eqref{eq:superdiff}, $g(u,v)$ is independent of $u\in U_x$. Averaging
against a convex representation of $v$ yields $g(u,v)=g(v,v)$ for
$u\in U_x$, hence $g(z,v)=g(v,v)$ for every $z\in\co U_x$. The reverse
Cauchy--Schwarz inequality gives
\[
 \normL z\normL v\le -g(z,v)=\normL v^2 ,
\]
so $\normL v=\vartheta(x)$ and $v=V(x)$ by Lemma~\ref{lem:V}. Finally,
$(b\circ x)'=\lambda g(v,v)=-\lambda\vartheta(x)^2$, while
$\normL{x'}=\lambda\vartheta(x)$.
\end{proof}

Thus the cone inclusion selects, almost everywhere, the maximal-norm
element of $\co U_x$. Although $C^b_x$ is a fan at singular points, the
cone condition must hold at almost every point of the curve, and the
comparison with the support functions of all the calibrated rays at once
leaves only the direction of $V$. In proper time the defect between the drop of $b$
and the elapsed proper time is
\begin{equation}\label{eq:defect}
 b(x(0))-b(x(t))-t
 =\int_0^t\bigl(\vartheta(x(s))-1\bigr)\,\dd s.
\end{equation}
The identity \eqref{eq:proper} holds only almost everywhere and does not
by itself allow one to differentiate $\vartheta\circ x$;
Sections~\ref{sec:support}--\ref{sec:invariance} provide the missing
control.

\section{Past invariance and global propagation}\label{sec:propagation}

This section proves Theorems~\ref{thm:invariance} and~\ref{thm:main}.
Sections~\ref{sec:support} and~\ref{sec:gronwall} provide the two
ingredients of the proof of past invariance: a support estimate whose
remainder is proportional to the strength defect $\vartheta-1$, and a
Gronwall lemma for forward averages. They are combined in
Section~\ref{sec:invariance}, global propagation is proved in
Section~\ref{sec:global}, and Section~\ref{sec:consequences} collects
consequences for the geometry of $\cS$.

\subsection{Angle-weighted support estimates}\label{sec:support}

For future unit timelike $u,w$ at one point put
$c(u,w)=-g(u,w)=\cosh\angle(u,w)$, where $\angle$ is the hyperbolic angle.

\begin{lemma}[Gauss-lemma cancellation]\label{lem:gauss}
Let $q$ range over a compact subset of $O$, let $u\in U_q$, and let $w$
range over a fixed compact family of future unit timelike vectors at $q$.
There are
$\sigma,\rho,K>0$, independent of $q,u,w$, such that for $0<r\le\rho$,
\begin{equation}\label{eq:taylor}
 d\bigl(\exp_q(\pm rw),\exp_q(\sigma u)\bigr)-\sigma
 \ge \mp r c(u,w)-Kr^2\bigl(c(u,w)^2-1\bigr).
\end{equation}
\end{lemma}

\begin{proof}
Choose $\sigma>0$ small and fixed. Compactness permits a uniform normal
coordinate ball $\norma{\xi}\le\eta$, inside a causally convex, convex
normal neighbourhood, on which $\exp_q\xi\ll\exp_q(\sigma u)$ and the
Lorentzian distance coincides with the smooth local distance. There is a
uniform bound $M$ for the second $\xi$-derivatives of
\[
 G_{q,u}(\xi)=d(\exp_q\xi,\exp_q(\sigma u))
                  -\sigma-g_q(u,\xi),
\]
by smooth dependence on $(q,u,\xi)$ on a compact subset of the smooth
distance domain, with $\sigma$ fixed away from zero.

Write $E_q=\exp_q$ and $\alpha(t)=E_q(tu)$. For sufficiently small $t$,
the connecting geodesic is maximizing and
$d(\alpha(t),\alpha(\sigma))=\sigma-t$. Moreover, for every
$\zeta\in T_qM$, first variation and the Gauss lemma give
\begin{align*}
 D_\xi[d(E_q\xi,E_q(\sigma u))]_{\xi=tu}[\zeta]
 &=g_{\alpha(t)}\bigl(\dot\alpha(t),(\dd E_q)_{tu}\zeta\bigr)\\
 &=g_q(u,\zeta).
\end{align*}
At $t=0$ the identity follows directly or by continuity. Consequently
\begin{equation}\label{eq:jet-vanish}
 G_{q,u}(tu)=0,\qquad DG_{q,u}(tu)=0.
\end{equation}
The differential vanishes in all directions, not only in the radial one;
no vanishing of the Hessian is asserted.

Set $c=c(u,w)$ and expand at $\pm rc\,u$, evaluating at $\pm rw$. Choose
$\rho$ so that both $r\norma{w}$ and $rc\norma{u}$ are at most $\eta/2$;
the whole straight Taylor segment then stays in the controlled ball. Since
\[
 w-cu\in u^\perp,\qquad g(w-cu,w-cu)=c^2-1,
\]
Taylor's theorem and \eqref{eq:jet-vanish} yield
\[
 G_{q,u}(\pm rw)\ge-\frac M2 r^2\norma{w-cu}^2
                         \ge-Kr^2(c^2-1),
\]
because the auxiliary norm and the positive definite metric
$g|_{u^\perp}$ are uniformly equivalent over the compact family of $u$'s.
This proves \eqref{eq:taylor}.
\end{proof}

\begin{lemma}[Initial-endpoint inequality]\label{lem:initial}
Let $q$ range over a compact subset of $O$ and $w$ over a fixed compact
family of future unit timelike vectors at $q$. There are $e\ge0$ and
$\rho>0$ such that, for $y=\exp_q(-rw)$ and $0<r\le\rho$,
\begin{equation}\label{eq:initial}
 b(q)-b(y)\ge
 r\left[1+(1-er)\bigl(\vartheta(q)-1\bigr)\right],
 \qquad er\le1.
\end{equation}
No maximizing property of $y$ is required.
\end{lemma}

\begin{proof}
Choose a finite convex representation $V(q)=\sum_i\alpha_iu_i$, with
$u_i\in U_q$. By \eqref{eq:support} and \eqref{eq:taylor} with the lower
sign,
\[
 b(q)-b(y)\ge d\bigl(y,\exp_q(\sigma u_i)\bigr)-\sigma
 \ge rc_i-Kr^2(c_i^2-1),\qquad c_i=c(u_i,w).
\]
Averaging gives $b(q)-b(y)\ge r\Theta-Kr^2\sum_i\alpha_i(c_i^2-1)$, with
$\Theta=\sum_i\alpha_ic_i=-g(V(q),w)\ge\vartheta(q)$ by the reverse
Cauchy--Schwarz inequality. Compactness gives $1\le c_i\le c_*$, so
$\sum_i\alpha_i(c_i^2-1)\le(c_*+1)(\Theta-1)$. With $e=K(c_*+1)$ and
$e\rho\le1$, the lower bound is $r[1+(1-er)(\Theta-1)]$, which proves
\eqref{eq:initial}.
\end{proof}

\begin{remark}
The crucial point in Lemmas~\ref{lem:gauss} and~\ref{lem:initial} is the
factor $c^2-1$ in \eqref{eq:taylor}. It turns the remainder into a
multiple of the strength defect $\vartheta-1$, rather than an additive
error independent of that defect. It rests on the vanishing of the value
and of the first differential in \eqref{eq:jet-vanish} along the entire
radial segment.
\end{remark}

\subsection{A forward-average Gronwall lemma}\label{sec:gronwall}

\begin{lemma}\label{lem:average}
Let $f:[0,L]\to[0,\infty)$ be bounded and upper semicontinuous. Suppose
that, for some $\omega\ge0$ and $h_0>0$,
\begin{equation}\label{eq:mean}
 \frac1h\int_t^{t+h}f(s)\,\dd s\ge(1-\omega h)f(t)
\end{equation}
whenever $0<h\le h_0$ and $0\le t<t+h\le L$. Then $e^{2\omega t}f(t)$ is
nondecreasing on $[0,L]$ and right-continuous on $[0,L)$. In particular,
\begin{equation}\label{eq:gronwall}
 f(t)\ge e^{-2\omega(t-s)}f(s)\qquad(0\le s\le t\le L).
\end{equation}
\end{lemma}

\begin{proof}
Let $\phi\in C_c^\infty((0,L))$ be nonnegative. Multiply \eqref{eq:mean} by
$\phi(t)$, integrate, subtract $\int f\phi$ and divide by $h$. By Fubini's
theorem and
\[
 \frac1h\int_{s-h}^s\phi(t)\,\dd t
 =\phi(s)-\frac h2\phi'(s)+O(h^2),
\]
letting $h\downarrow0$ gives $-\frac12\int f\phi'\ge-\omega\int f\phi$.
Thus $Df+2\omega f\,\dd t\ge0$ in the sense of distributions, and
$e^{2\omega t}f(t)$ agrees almost everywhere on $(0,L)$ with a
nondecreasing right-continuous function $G$, extended to $0$ by its right
limit.

The point values are recovered as follows. At every $t\in[0,L)$, by
monotonicity of $G$ and almost-everywhere agreement,
\[
 \lim_{h\downarrow0}\frac1h\int_t^{t+h}f(s)\,\dd s=e^{-2\omega t}G(t),
\]
so \eqref{eq:mean} gives $f(t)\le e^{-2\omega t}G(t)$. Conversely, upper
semicontinuity along a sequence of points of almost-everywhere agreement
decreasing to $t$ gives the reverse inequality. Hence
$e^{2\omega t}f(t)=G(t)$ for every $t<L$. At $L$, upper semicontinuity
gives $e^{2\omega L}f(L)\ge G(L-)$, which is sufficient for monotonicity on
the closed interval.
\end{proof}

\subsection{Proof of Theorem~\ref{thm:invariance}}\label{sec:invariance}

\begin{proof}[Proof of Theorem~\ref{thm:invariance}]
All constants are uniform on a compact neighborhood of $x([0,L])$. For
uniformly small $h>0$, the subarc $x([t,t+h])$ lies in a causally convex,
convex normal neighborhood in which its endpoints are joined by the
maximizing timelike geodesic. Write
\[
 q=x(t),\qquad y=x(t+h)=\exp_q(-rw),\qquad \normL w=1,
\]
with $w$ future directed and $r=d(y,q)$. Since the subarc has proper
length $h$, $r\ge h$. Since the auxiliary speed of $x$ is bounded,
$\norma{\exp_q^{-1}(y)}\le B_1h$, and therefore $r\le\Lambda_1h$ and
$\norma{w}\le B_1$, for constants $B_1,\Lambda_1$. Unit timelike vectors of
bounded auxiliary norm form a compact family, so Lemma~\ref{lem:initial}
applies uniformly.

Put $f(t)=\vartheta(x(t))-1$. By \eqref{eq:defect},
\begin{equation}\label{eq:drop}
 b(x(t))-b(x(t+h))=h+\int_t^{t+h}f(s)\,\dd s.
\end{equation}
On the other hand, \eqref{eq:initial} with $r\ge h$ and $er\le e\Lambda_1h$
gives, for $e\Lambda_1h\le1$,
\[
 b(x(t))-b(x(t+h))\ge r\bigl[1+(1-er)f(t)\bigr]
 \ge h\bigl[1+(1-e\Lambda_1h)f(t)\bigr].
\]
Comparison with \eqref{eq:drop} proves \eqref{eq:mean} with
$\omega=e\Lambda_1$. The function $f$ is bounded, nonnegative and upper
semicontinuous by Lemma~\ref{lem:strength}, and Lemma~\ref{lem:average}
yields \eqref{eq:main} with $E=2e\Lambda_1$. The last assertions follow
from Lemma~\ref{lem:strength}, applied to compact subarcs of locally
absolutely continuous curves after reparametrization.
\end{proof}

\begin{corollary}[Right derivative along cone-causal curves]\label{cor:rightderiv}
Let $x$ be a past-directed $C^b$-causal curve parametrized by proper time.
Then $\vartheta\circ x$ is locally of bounded variation and right-continuous
at nonterminal parameters, $x$ has a right derivative at every nonterminal
parameter,
\[
 \frac{\dd^+x}{\dd t}(t)=-\frac{V(x(t))}{\vartheta(x(t))},
\]
and this right derivative is right-continuous.
\end{corollary}

\begin{proof}
By the proof of Theorem~\ref{thm:invariance},
$e^{Et}(\vartheta(x(t))-1)$ is nondecreasing and right-continuous. For
$t_j\downarrow t$, any limit of $V(x(t_j))$ belongs to $\co U_{x(t)}$, by
the closed graph property, and has norm $\vartheta(x(t))$, by
right-continuity; hence it is $V(x(t))$ by Lemma~\ref{lem:V}. Thus
$V\circ x$ is right-continuous in local trivializations, and integrating
\eqref{eq:proper} gives the right derivative.
\end{proof}

\subsection{Proof of Theorem~\ref{thm:main}}\label{sec:global}

\begin{proof}[Proof of Theorem~\ref{thm:main}]
Let $p\in\cS$. By Remark~\ref{rem:cones}, $x\mapsto-C^b_x$ is a closed cone
structure on $O$, so \cite[Theorem~2.1]{minguzzi17} provides a continuous
causal curve for it starting at $p$ and inextendible in $O$, namely a
past-directed $C^b$-causal curve $x$ starting at $p$ which is past
inextendible in $O$. It is past inextendible in $M$: a past endpoint $q$
would lie in $J^-(p)\subset O$, by the closedness of $J^-$ in globally
hyperbolic spacetimes and $J^-(O)\subset O$, and the curve could then be
extended in $O$. Every past-inextendible causal curve meets every level of
the Cauchy temporal function $T$ below $T(p)$, and $T$ is strictly
decreasing along it, so it can be parametrized as
$\gamma:(-\infty,T(p)]\to O$ with $T(\gamma(\tau))=\tau$; then
\eqref{eq:main-inclusion} holds by definition of $\cF$. By
Lemma~\ref{lem:F}, $\gamma$ is locally Lipschitz and locally uniformly
timelike. Finally, Theorem~\ref{thm:invariance}, applied to the compact
subarcs $\gamma([a,T(p)])$ in proper time, gives
$\vartheta(\gamma(\tau))-1\ge e^{-E\ell}(\vartheta(p)-1)>0$ for
$\tau\in[a,T(p)]$, where $\ell$ is the proper length of that subarc. The
same argument applies to every past-inextendible past-directed
$C^b$-causal curve starting at $p$.
\end{proof}

\begin{remark}[The inclusion $x'\in-\co U_x$]\label{rem:inclusion}
By Lemma~\ref{lem:rigidity}, in proper time $x'=-V(x)/\vartheta(x)$. The
parameter
\[
 \varsigma(t)=\int_0^t\frac{\dd a}{\vartheta(x(a))},
\]
which satisfies $\dd t/\dd\varsigma=\vartheta(x)$ and is a locally
bi-Lipschitz change of parameter, since $\vartheta\ge1$ is upper
semicontinuous and hence locally bounded, gives
\[
 \frac{\dd x}{\dd\varsigma}=-V(x)\in-\co U_x
 \quad\text{almost everywhere},\qquad
 \frac{\dd}{\dd\varsigma}\,b(x)=-\vartheta(x)^2\le-1 .
\]
This is the inclusion proposed in the paragraph {\em Expected global
propagation} after \cite[Theorem~5.21]{minguzzi26e}, with the selection of the
maximal-norm element forced by the cone condition.
\end{remark}

\subsection{Consequences for the geometry of the singular set}\label{sec:consequences}

\begin{corollary}\label{cor:geometry}
Under the standing assumptions:
\begin{enumerate}
\item For $p\in\cS$ and every Cauchy hypersurface $\Sigma$ with
      $p\in J^+(\Sigma)$,
      $\cS\cap\Sigma\cap J^-(p)\ne\varnothing$, while
      $\mathcal H^n(\cS\cap\Sigma)=0$.
\item If $\cS\ne\varnothing$, then $T(\cS)$ is an interval unbounded
      below. Every connected component of $\cS$ is unbounded below for
      $T$, so no nonempty connected component is relatively compact in
      $M$.
\item If $b$ is differentiable at every point of $\Sigma\cap O$ for a
      Cauchy hypersurface $\Sigma$, it is differentiable on
      $J^+(\Sigma)\cap O$ and is $C^{1,1}_{\mathrm{loc}}$ on
      $I^+(\Sigma)\cap O$.
\end{enumerate}
\end{corollary}

\begin{proof}
For (1), the case $p\in\Sigma$ is immediate. Otherwise $p\in I^+(\Sigma)$.
Extend the curve of Theorem~\ref{thm:main} to a future-inextendible timelike
curve. Its unique intersection with $\Sigma$ must occur on the original
past portion, since the future extension lies in $I^+(\Sigma)$. The set
$\cS\cap\Sigma$ is $\mathcal H^n$-null because $\Sigma$ is achronal
\cite[Theorem~5.21(c)]{minguzzi26e}.
For (2), the curve through every $p\in\cS$ is connected, stays in its
connected component, and has $T$-image $(-\infty,T(p)]$. The union of
these intervals is an interval unbounded below. A relatively compact
subset of $M$ has bounded $T$-image.
For (3), (1) excludes singular points in $J^+(\Sigma)\cap O$.
Semiconcavity gives $C^1$ regularity on the open future region, and the
$C^{1,1}_{\mathrm{loc}}$ conclusion follows from
\cite[Theorem~5.21(b)]{minguzzi26e}, applied to an open set consisting of
differentiability points.
\end{proof}

\begin{remark}[Strength and angular diameter]\label{rem:diam}
Let $D_x$ be the diameter of $U_x$ for the hyperbolic distance $\angle$ on
the future unit hyperboloid of $T_xM$. The midpoint of a diametral pair has
Lorentzian norm $\cosh(D_x/2)$, and every $v=\sum_i\lambda_iu_i\in\co U_x$
satisfies $\normL{v}^2=\sum_{i,j}\lambda_i\lambda_j\cosh\angle(u_i,u_j)\le
\cosh D_x$. Hence
\[
 \cosh(D_x/2)\le\vartheta(x)\le\sqrt{\cosh D_x},
\]
so that $\vartheta(x)-1$ and $D_x^2$ are comparable on compact subsets of
$O$, where $D_x$ is also comparable with the diameter of $U_x$ for any
auxiliary Riemannian metric. Therefore \eqref{eq:main} says that the
diameter of $U_x$ cannot decay faster than exponentially in proper time
along any past-directed $C^b$-causal curve, and the last statement of
Theorem~\ref{thm:main} says that it is bounded below on every compact
temporal interval. This is the global and timelike counterpart of
\cite[Theorem~5.21(d)]{minguzzi26e}, which provides a Lipschitz singular curve
$\sigma:[0,\rho]\to\cS$ along which the diameter of $U_{\sigma(s)}$ is
bounded below.
\end{remark}

\section{Past uniqueness and the past semi-flow}\label{sec:unique}

This section proves Theorem~\ref{thm:unique}. Part~(1) uses
Lemma~\ref{lem:rigidity}, which puts all proper-time cone-causal curves
into one autonomous gradient inclusion, and a local convexification; it
uses neither Theorem~\ref{thm:invariance} nor
Corollary~\ref{cor:rightderiv}. The order comparison is related to the
Lorentzian gradient-flow framework discussed in
\cite[Section~2.3]{Almost}; no external gradient-flow uniqueness theorem
is invoked.

Throughout the section put $h=-g$, of signature $(+,-,\ldots,-)$, reverse
the time orientation, and set $f=-b$. Thus $h$-future means $g$-past; the
Levi-Civita connection and the norm $\normL\cdot$ are unchanged, and
$\ll_h$, $\le_h$ denote the chronological and causal relations for the
reversed time orientation. Subdifferentials are taken in local
coordinates: $D^-f(x)$ is the set of covectors $q$ such that
$f(y)\ge f(x)+q(y-x)+o(|y-x|)$ as $y\to x$. Since $D^-f=-D^+b$,
\eqref{eq:superdiff} becomes
\begin{equation}\label{eq:sub}
 D^-f(x)=\{g_x(v,\cdot):v\in\co U_x\},
\end{equation}
and for $F=e^{kf}$, $k>0$, the chain rule gives $D^-F=ke^{kf}D^-f$.

\subsection{Local causal convexification}

\begin{lemma}\label{lem:convexify}
Every point of $O$ has a relatively compact, causally convex, convex normal
neighbourhood $N\subset O$ with the following property: for some $k>0$,
the function
\[
 F=e^{kf}=e^{-kb}
\]
is convex along every affinely parametrized $h$-causal geodesic contained
in $N$, and for every such geodesic $\sigma:[0,1]\to N$ from $a$ to $c$
and all $q_a\in D^-F(a)$, $q_c\in D^-F(c)$,
\begin{equation}\label{eq:causal-support}
 q_a(\dot\sigma(0))\le F(c)-F(a)\le q_c(\dot\sigma(1)).
\end{equation}
\end{lemma}

\begin{proof}
Fix $x_0\in O$, a chart around it and coordinate balls
$B'\Subset B\Subset O$ centred at $x_0$. Let $\mathcal C_x$ be the closed
future $h$-causal cone at $x$, namely the closed past $g$-causal cone, and
compute pairings of vectors at different points in coordinates.

{\em Uniform steepness.} By the proof of Lemma~\ref{lem:F}, the vectors of
$\co U_y$, $y\in\overline B$, form a compact family of future $g$-timelike
vectors, and a future $g$-timelike vector pairs positively under $g$ with
every nonzero past $g$-causal vector. A compactness argument, using the
continuity of the cone field, gives $c,\eta>0$ such that
\begin{equation}\label{eq:steep}
 g_y(v,w)\ge c\norma w\qquad
 \bigl(y\in\overline B,\ v\in\co U_y,\ x\in\overline{B'},\
 |x-y|\le\eta,\ w\in\mathcal C_x\bigr).
\end{equation}
Indeed, otherwise there would be such data with $\norma{w_j}=1$,
$|x_j-y_j|\to0$ and $g_{y_j}(v_j,w_j)\to\ell\le0$; a subsequence would
converge to $y\in\overline{B'}$, $v\in\co U_y$ and a unit $w\in\mathcal C_y$
with $g_y(v,w)\le0$, which is impossible.

{\em Mollification.} Adding \eqref{eq:semiconc} at $y=x\pm\xi$ gives
$b(x+\xi)+b(x-\xi)-2b(x)\le2A|\xi|^2$; hence the coordinate Hessian of $f$
satisfies $D^2f\ge-2A\,I$ in the sense of distributions on $B$. Let
$f_\epsilon$, $0<\epsilon\le\eta$, be standard mollifications of $f$,
defined on $\overline{B'}$ for small $\epsilon$. Then
$D^2f_\epsilon\ge-2A\,I$, the differentials $\dd f_\epsilon$ are bounded by
a Lipschitz constant of $f$ on $B$, and $f_\epsilon\to f$ uniformly on
$\overline{B'}$. With bounded Christoffel symbols, there is $C_0\ge0$ such
that on $\overline{B'}$
\[
 \operatorname{Hess}_hf_\epsilon(w,w)
 =D^2f_\epsilon(w,w)-\Gamma^\lambda_{\mu\nu}w^\mu w^\nu\partial_\lambda f_\epsilon
 \ge-C_0\norma w^2 .
\]
Moreover, $\dd f_\epsilon(x)=\int\dd f(x-z)\rho_\epsilon(z)\,\dd z$ is an
average of differentials $\dd f(y)$ at differentiability points $y$ with
$|x-y|\le\epsilon$, and $\dd f(y)=g_y(u,\cdot)$ with
$u=-\nabla b(y)\in U_y$ by fact~4. By \eqref{eq:steep},
\[
 \dd f_\epsilon(x)(w)\ge c\norma w\qquad
 (x\in\overline{B'},\ w\in\mathcal C_x).
\]

{\em Convexity.} Choose $k>0$ with $kc^2\ge C_0$. For every $h$-causal
vector $w$ at a point of $\overline{B'}$, applying the last inequality to
$w$ or to $-w$,
\[
 \operatorname{Hess}_h(e^{kf_\epsilon})(w,w)
 =ke^{kf_\epsilon}\bigl(\operatorname{Hess}_hf_\epsilon(w,w)
   +k[\dd f_\epsilon(w)]^2\bigr)
 \ge ke^{kf_\epsilon}(kc^2-C_0)\norma w^2\ge0 .
\]
Let $N\subset B'$ be a relatively compact, causally convex, convex normal
neighbourhood of $x_0$. For an affinely parametrized $h$-causal geodesic
$\sigma$ in $N$,
$(e^{kf_\epsilon}\circ\sigma)''
=\operatorname{Hess}_h(e^{kf_\epsilon})(\dot\sigma,\dot\sigma)\ge0$, so
$e^{kf_\epsilon}\circ\sigma$ is convex, and so is its uniform limit
$F\circ\sigma$.

{\em Support inequality.} If $q_a\in D^-F(a)$, then
$(F\circ\sigma)'_+(0)\ge q_a(\dot\sigma(0))$, and convexity gives
$F(c)-F(a)\ge(F\circ\sigma)'_+(0)$, which is the first inequality in
\eqref{eq:causal-support}. Applying this to the reversed geodesic
$t\mapsto\sigma(1-t)$ gives $F(a)-F(c)\ge q_c(-\dot\sigma(1))$, the second
inequality.
\end{proof}

\subsection{A common parameter and chronological comparison}

Let $N$ and $k$ be as in Lemma~\ref{lem:convexify}, and let $x$ be a
past-directed $C^b$-causal curve in $N$, parametrized by proper time $r$
starting at $r=0$. Since $\vartheta\circ x$ is Borel, at least one and
locally bounded, we may define
\begin{equation}\label{eq:parameter}
 s(r)=\int_0^r\frac{\dd a}{ke^{-kb(x(a))}\vartheta(x(a))}.
\end{equation}
The integrand is bounded above and away from zero on compact subintervals,
so this is a locally bi-Lipschitz change of parameter. For
$\gamma(s)=x(r(s))$, \eqref{eq:proper} gives
\begin{equation}\label{eq:flow}
 \gamma'=-ke^{-kb(\gamma)}V(\gamma),\qquad
 h(\gamma',\cdot)\in D^-F(\gamma)\qquad\text{a.e.},
\end{equation}
the membership because $h(\gamma',\cdot)=ke^{-kb(\gamma)}g(V(\gamma),\cdot)$,
by \eqref{eq:sub} and $D^-F=ke^{kf}D^-f$. Conversely, if a locally
Lipschitz curve satisfies the inclusion in \eqref{eq:flow}, then
$\gamma'=-ke^{-kb(\gamma)}v$ with $v\in\co U_\gamma$ almost everywhere, so
it is past-directed $C^b$-causal and, by Lemma~\ref{lem:rigidity}, it also
satisfies the equation in \eqref{eq:flow}. The inclusion is autonomous:
time translates of solutions are solutions. Its solutions are locally
uniformly $h$-timelike and future directed for $h$.

\begin{lemma}[Preservation of strict chronological order]\label{lem:order}
Let $N$ and $k$ be as in Lemma~\ref{lem:convexify}, and let
$\alpha,\beta:[0,S]\to N$ be Lipschitz curves satisfying the inclusion in
\eqref{eq:flow}. If $\alpha(0)\ll_h\beta(0)$, then
$\alpha(s)\ll_h\beta(s)$ for every $s\in[0,S]$, and the $h$-Lorentzian
distance in $N$ from $\alpha(s)$ to $\beta(s)$ is nondecreasing in $s$.
\end{lemma}

\begin{proof}
On $N$ use the smooth function
\[
 \Omega(a,c)=\tfrac12h_a(\exp_a^{-1}c,\exp_a^{-1}c).
\]
Since $N$ is causally convex and convex normal, $a\ll_hc$ holds if and only
if $\exp_a^{-1}c$ is future $h$-timelike, and then $\Omega(a,c)$ is half
the square of the $h$-Lorentzian distance in $N$ from $a$ to $c$. Let
$\sigma_s:[0,1]\to N$ be the affine geodesic from $\alpha(s)$ to
$\beta(s)$. The function $\omega(s)=\Omega(\alpha(s),\beta(s))$ is
Lipschitz and, by the first variation formula for the energy of geodesics
with moving endpoints, at almost every $s$ with $\alpha(s)\ll_h\beta(s)$,
\[
 \omega'(s)=h\bigl(\beta'(s),\dot\sigma_s(1)\bigr)
           -h\bigl(\alpha'(s),\dot\sigma_s(0)\bigr)
 =q_\beta(\dot\sigma_s(1))-q_\alpha(\dot\sigma_s(0))\ge0,
\]
where $q_\alpha=h(\alpha'(s),\cdot)\in D^-F(\alpha(s))$,
$q_\beta=h(\beta'(s),\cdot)\in D^-F(\beta(s))$, and the inequality is
\eqref{eq:causal-support}. Let $J$ be the connected component of
$\{s\in[0,S]:\alpha(s)\ll_h\beta(s)\}$ containing $0$; it is an interval,
relatively open in $[0,S]$, on which $\omega\ge\omega(0)>0$. It is also
closed: at a limit point $s_*$ of $J$, $\exp^{-1}_{\alpha(s_*)}\beta(s_*)$
is a limit of future $h$-timelike vectors with $h$-square at least
$2\omega(0)>0$, hence it is future $h$-timelike, and $s_*\in J$. Thus
$J=[0,S]$.
\end{proof}

\subsection{Proof of Theorem~\ref{thm:unique}(1)}

Let $x_1,x_2$ be past-directed $C^b$-causal curves parametrized by proper
time, with $x_1(0)=x_2(0)=p$. Choose $N$ and $k$ for the point $p$ by
Lemma~\ref{lem:convexify}, and reparametrize the initial portions of $x_1$
and $x_2$ contained in $N$ by \eqref{eq:parameter}. This gives curves
$\alpha,\beta$ satisfying \eqref{eq:flow}, with $\alpha(0)=\beta(0)=p$,
which we restrict to $[0,S+\epsilon_0]$, $S,\epsilon_0>0$, so that they lie
in $N$. Since $\beta$ is locally uniformly timelike,
$p\ll_h\beta(\epsilon)$ for $0<\epsilon\le\epsilon_0$. The curve
$s\mapsto\beta(s+\epsilon)$, $s\in[0,S]$, satisfies the same autonomous
inclusion, so Lemma~\ref{lem:order} gives
\[
 \alpha(s)\ll_h\beta(s+\epsilon)\qquad(s\in[0,S]).
\]
Letting $\epsilon\downarrow0$ and using the closedness of the causal
relation of $N$ gives $\alpha(s)\le_h\beta(s)$. Exchanging the roles of the
curves gives $\beta(s)\le_h\alpha(s)$, hence $\alpha(s)=\beta(s)$, as $N$
contains no closed causal curves.

Thus the two curves have the same image near $p$, traversed in the same
direction. Proper time along this common curve, and likewise the drop of
$T$, is determined by the curve, so $x_1=x_2$ near $r=0$. Let $r_*$ be the
supremum of the $r$ such that $x_1=x_2$ on $[0,r]$. By continuity
$x_1=x_2$ on $[0,r_*]$, and if $r_*$ were smaller than the endpoint of the
common domain, the local argument applied to $r\mapsto x_i(r_*+r)$ at the
point $x_1(r_*)$ would contradict the definition of $r_*$. The statement
for the $T$-parametrizations follows, since each of them is related to the
proper-time parametrization by an increasing bijection determined by the
curve. \qed

\subsection{Proof of Theorem~\ref{thm:unique}(2)}

{\em Existence, uniqueness and the semigroup law.} The proof of
Theorem~\ref{thm:main} uses $p\in\cS$ only in its final strength estimate.
Thus \cite[Theorem~2.1]{minguzzi17} gives, for every $p\in O$, a
past-directed $C^b$-causal curve starting at $p$ and past inextendible in
$M$, which can be parametrized by the drop of $T$ over $[0,\infty)$. By
part~(1) it is unique, and we call it $s\mapsto\Psi_s(p)$. For $t\ge0$ the
curve $s\mapsto\Psi_{s+t}(p)$ is past-directed $C^b$-causal, starts at
$\Psi_t(p)$ and satisfies $T(\Psi_{s+t}(p))=T(\Psi_t(p))-s$; by uniqueness
it is $s\mapsto\Psi_s(\Psi_t(p))$. The inclusion $\Psi_s(\cS)\subset\cS$ is
Theorem~\ref{thm:invariance}, applied in proper time, and the independence
of the oriented trajectories from $T$ is part~(1).

{\em The right derivative.} Let $x$ be the proper-time parametrization of
$s\mapsto\Psi_s(p)$. By Corollary~\ref{cor:rightderiv} and its proof, $x$
has the right derivative $-V(x)/\vartheta(x)$ at every parameter, and
$V\circ x$ and $\vartheta\circ x$ are right-continuous. Hence the drop
$\mu(r)=T(p)-T(x(r))$ is strictly increasing, continuous, and has at
every $r$ the positive right-continuous right derivative
$\dd T(V(x))/\vartheta(x)$. Its inverse $r(s)$ therefore has the right
derivative $\vartheta(x)/\dd T(V(x))$ at every $s$, and the chain rule for
right derivatives gives \eqref{eq:generator}, which is right-continuous in
$s$ since $V\circ x$ is. In particular
$\frac{\dd}{\dd s}\Psi_s(p)=-V/\dd T(V)$ at almost every $s$, consistently
with \eqref{eq:proper}.

{\em Continuity.} Let $p_j\to p$ in $O$ and $S>0$. Choose
$q\in I^+(p)\cap O$, which exists since $O$ is open; for large $j$,
$p_j\ll q$ and $T(p_j)\ge T(p)-1$. The curves $\Psi_s(p_j)$, $s\in[0,S]$,
lie in
\[
 K=J^-(q)\cap T^{-1}\bigl([T(p)-S-1,T(q)]\bigr),
\]
which is compact by global hyperbolicity and contained in $O$ by fact~1.
Their velocities belong almost everywhere to $-\cF$, which is uniformly
bounded over $K$ by Lemma~\ref{lem:F}, so the curves are equi-Lipschitz. By
the Arzel\`a--Ascoli theorem, every subsequence has a further subsequence
converging uniformly on $[0,S]$ to a curve $\gamma$ with $\gamma(0)=p$ and
$T(\gamma(s))=T(p)-s$. The argument of the paragraph {\em The limiting
inclusion} in the proof of Proposition~\ref{prop:chains}, applied to the
exact velocities instead of the approximate ones, shows that
$\dot\gamma\in-\cF_\gamma$ almost everywhere, so that $\gamma$ is a
past-directed $C^b$-causal curve. By part~(1), $\gamma(s)=\Psi_s(p)$.
Therefore the whole sequence converges uniformly on $[0,S]$, and, with the
uniform Lipschitz bound in $s$, $\Psi$ is jointly continuous. Its joint
local Lipschitz continuity is proved in Section~\ref{sec:lip}. \qed

\subsection{Joint local Lipschitz regularity}\label{sec:lip}

We prove that $\Psi$ is jointly locally Lipschitz, which completes the
proof of Theorem~\ref{thm:unique}(2). Lipschitz continuity refers to
coordinate charts, equivalently to the auxiliary metric, and to the
$T$-normalization; the local constants are not uniform in $s\in[0,\infty)$.

{\em The autonomous local semi-flow.} Fix $x_0\in O$, and let $N$ and $k$ be as
in Lemma~\ref{lem:convexify}, with $h=-g$ and the reversed time
orientation as in Section~\ref{sec:unique}. Put
$X(x)=-ke^{-kb(x)}V(x)$. For $p\in N$ let $t\mapsto\Gamma_t(p)$ be the
trajectory of $\Psi$ from $p$ in the parameter \eqref{eq:parameter}, so that
$\dd\Gamma_t(p)/\dd t=X(\Gamma_t(p))$ almost everywhere, as long as
$\Gamma_t(p)\in N$. By Theorem~\ref{thm:unique}(1) and the autonomy of
\eqref{eq:flow}, $\Gamma_{t+t'}=\Gamma_t\circ\Gamma_{t'}$. No continuity of
$X$ is used.

{\em Coordinates and constants.} Choose coordinates
$z=(z^0,\mathbf z)$ around $x_0$ in which $h_{x_0}$ is the Minkowski form
of signature $(+,-,\ldots,-)$ and $\partial_0$ is future directed for $h$,
and let $|\cdot|$ be the Euclidean coordinate norm. The vectors $X(x)$
belong, for $x$ near $x_0$, to an arbitrarily small neighbourhood of the
compact set $-ke^{-kb(x_0)}\co U_{x_0}$ of future $h$-timelike vectors, by
the upper semicontinuity of $x\mapsto\co U_x$ (Lemma~\ref{lem:F}), and the
$h$-cones vary continuously. Hence there are constants $0<\kappa<\kappa'<1$
and $m,L,B>0$, and a neighbourhood $W\subset N$ of $x_0$, such that at every
$x\in W$
\begin{equation}\label{eq:lipbounds}
 X^0(x)\ge m,\qquad |\mathbf X(x)|\le\kappa X^0(x),\qquad |X(x)|\le L,
\end{equation}
every nonzero $w$ with $w^0>0$ and $|\mathbf w|\le\kappa'w^0$ is future
$h$-timelike, and every future $h$-causal $w$ satisfies
$|\mathbf w|\le Bw^0$. Let $N'\subset W$ be a causally convex, convex normal
neighbourhood of $x_0$, let $W_1\subset N'$ be a coordinate ball of radius
$\rho_1$ centred at $x_0$, and let $W_0$ be the concentric ball of radius
$\rho_1/2$. With $t_0=\rho_1/(4L)$, the points $\Gamma_t(p)$,
$p\in W_0$, $0\le t\le2t_0$, lie in $W_1$, which is convex in coordinates.
Since $N'\subset N$ is causally convex and convex normal, Lemma~\ref{lem:order}
holds with $N'$ in place of $N$.

{\em Lipschitz estimate in the semi-flow parameter.} Let $p,q\in W_0$,
$\delta=|p-q|>0$, and $\epsilon=C\delta$, with
\begin{equation}\label{eq:lipC}
 C>\frac{1+\kappa'}{m(\kappa'-\kappa)} .
\end{equation}
If $\epsilon>t_0$, then
$|\Gamma_t(p)-\Gamma_t(q)|\le2\rho_1\le(2\rho_1C/t_0)\delta=8LC|p-q|$
for $t\le t_0$. Let $\epsilon\le t_0$. The displacement
$\Gamma_\epsilon(q)-q=\int_0^\epsilon X(\Gamma_u(q))\,\dd u$ has time
component at least $m\epsilon$, and its spatial part has norm at most
$\kappa$ times its time component, by \eqref{eq:lipbounds} and the
convexity of the cone of slope $\kappa$. Hence $d=\Gamma_\epsilon(q)-p$
satisfies
\[
 d^0\ge m\epsilon-\delta>0,\qquad
 |\mathbf d|\le\kappa(d^0+\delta)+\delta<\kappa'd^0,
\]
the last inequality by \eqref{eq:lipC}. The straight segment from $p$ to
$\Gamma_\epsilon(q)$ lies in $W_1$ and has tangent in the cone of slope
$\kappa'$, so it is future $h$-timelike, and $p\ll_h\Gamma_\epsilon(q)$.
Likewise $q\ll_h\Gamma_\epsilon(p)$. Lemma~\ref{lem:order}, applied to the
curves $t\mapsto\Gamma_t(p)$ and $t\mapsto\Gamma_{t+\epsilon}(q)$ and to
the pair with $p$ and $q$ exchanged, gives for $0\le t\le t_0$
\[
 p_t\ll_hq_t^+,\qquad q_t\ll_hp_t^+,\qquad
 p_t=\Gamma_t(p),\ q_t=\Gamma_t(q),\ p_t^+=\Gamma_{t+\epsilon}(p),\
 q_t^+=\Gamma_{t+\epsilon}(q).
\]
Moreover $|p_t^+-p_t|\le L\epsilon$ and $|q_t^+-q_t|\le L\epsilon$. Since
$z^0$ increases along future $h$-causal curves in $N'$, the two relations
give $p_t^0<q_t^0+L\epsilon$ and $q_t^0<p_t^0+L\epsilon$, so that
$|p_t^0-q_t^0|\le L\epsilon$. Integrating $|\mathbf w|\le Bw^0$ along a
future $h$-causal curve from $p_t$ to $q_t^+$ in $N'$ gives
$|\mathbf p_t-\mathbf q_t^+|\le B(q_t^{+0}-p_t^0)\le2BL\epsilon$, hence
$|\mathbf p_t-\mathbf q_t|\le(2B+1)L\epsilon$ and
\[
 |\Gamma_t(p)-\Gamma_t(q)|\le(2B+2)LC\,|p-q|\qquad(0\le t\le t_0).
\]
In both cases
$|\Gamma_t(p)-\Gamma_t(q)|\le\max\{8,2B+2\}LC\,|p-q|$ for
$p,q\in W_0$ and $0\le t\le t_0$, and with $|X|\le L$, $\Gamma$ is
Lipschitz on $[0,t_0]\times W_0$.

{\em Temporal normalization.} For $p\in W_0$ and $t\in[0,t_0]$ put
$H(t,p)=T(p)-T(\Gamma_t(p))$, a Lipschitz function, with constant $C_H$ in
$p$. The vectors $X$ are past directed for $g$ and lie in a compact subset
of the timelike cone bundle over $\overline W$, so $-\dd T(X)\ge m_T>0$
there, and $H(t_2,p)-H(t_1,p)\ge m_T(t_2-t_1)$ for $t_1\le t_2$. For
$0\le s\le s_1=m_Tt_0$ let $\tau(s,p)$ be defined by $H(\tau(s,p),p)=s$.
Then $\Psi_s(p)=\Gamma_{\tau(s,p)}(p)$, and
\[
 m_T|\tau(s,p)-\tau(r,q)|\le|H(\tau(s,p),q)-H(\tau(r,q),q)|
 \le C_H|p-q|+|s-r|,
\]
as $H(\tau(s,p),p)=s$ and $H(\tau(r,q),q)=r$. Hence $\Psi$ is Lipschitz on
$[0,s_1]\times W_0$.

{\em Finite times.} Let $p\in O$ and $S>0$. Cover the compact arc
$\Psi_{[0,S]}(p)$ by finitely many neighbourhoods of the type $W_0$, and
choose $0=\sigma_0<\sigma_1<\dots<\sigma_N=S$ such that each arc
$\Psi_{[\sigma_i,\sigma_{i+1}]}(p)$ starts in one of them, $W_0^{(i)}$
say, with $\sigma_{i+1}-\sigma_i$ smaller than the corresponding time
$s_1^{(i)}$. By the continuity of $\Psi$, for $q$ near $p$ the points
$\Psi_{\sigma_i}(q)$ belong to $W_0^{(i)}$, and the semigroup law writes
$\Psi_s$, $0\le s\le S$, as a composition of at most $N$ Lipschitz maps.
Hence $\Psi$ is Lipschitz on $[0,S]\times W'$ for a neighbourhood $W'$ of
$p$.

\begin{remark}[Merging, and the future direction]\label{rem:merge}
Past uniqueness does not imply injectivity of $\Psi_s$. In $1+1$ Minkowski
space, with $g=-\dd t^2+\dd z^2$, set
\[
 b(t,z)=At-a|z|,\qquad a>0,\quad A=\sqrt{1+a^2},
\]
the minimum of two affine functions with past unit timelike gradients. Off
the axis $z=0$ the only calibrated direction is $(A,a\operatorname{sgn}z)$,
while on the axis both $(A,\pm a)$ occur, so that $V=(A,0)$ and
$\vartheta=A$. With $T=t$ and $s=t(0)-t$, the trajectories satisfy
$\dd z/\dd s=-(a/A)\operatorname{sgn}z$ off the axis and $\dd z/\dd s=0$ on
it: they reach the singular axis in finite time and coincide afterwards.
A past-directed curve which left the axis would enter a region where the
only past direction points back to it; this is how the wide cones on the
axis are compatible with past uniqueness. In the future direction, from a point of the axis, the two calibrated rays
and the curve along the axis are all future-directed $C^b$-causal, so
future uniqueness fails at singular points. This is a local affine model,
not an asserted global example with compact Cauchy hypersurfaces.
\end{remark}

\section{A constructive proof by maximization on past spheres}\label{sec:spheres}

The curves of Theorem~\ref{thm:main} can also be obtained, without the
existence theorem for closed cone structures and without
Theorem~\ref{thm:invariance}, as limits of chains of maximization steps.
By Theorem~\ref{thm:unique}(1), whose proof uses neither of these results,
the limit does not depend on the choices made along the chains.

Fix a compact $C\subset O$ and a compact neighborhood
$C'\subset O$ with $C\subset\operatorname{int}C'$. Choose $\delta>0$ so
that $\overline{B_a(x,\delta)}\subset C'$ for $x\in C$ and all the small
geodesics and distance functions below are controlled in normal
neighborhoods. In detail, take finite nested normal and causally convex
neighborhoods; for sufficiently close pairs, the Lorentzian distance agrees
with the local normal-neighborhood distance and is smooth on the
chronologically related pairs used below. Coordinate norms,
exponential-map derivatives and local semiconcavity constants are then
uniformly controlled. The normalized vectors
\[
 W_C=\left\{\frac{v}{\normL{v}}:v\in\co U_q,\ q\in C'\right\}
\]
form a compact family of future unit timelike vectors. For $x\in C$ set
\[
 Q_r(x)=\{q\in J^-(x)\cap\overline{B_a(x,\delta)}:d(q,x)\ge r\}.
\]

\begin{lemma}[Sphere step]\label{lem:step}
There are $r_C,\Lambda_C>0$ such that for $x\in C$ and $0<r\le r_C$,
$b|_{Q_r(x)}$ attains its maximum. Every maximizing point $z$ satisfies
\[
 z\ll x,\qquad d(z,x)=r,\qquad d_a(z,x)\le\Lambda_C r,
\]
and the connecting geodesic has auxiliary length at most $\Lambda_Cr$. If
$w_z$ is its future unit tangent at $z$, then
\begin{equation}\label{eq:multiplier}
 \nu w_z\in\co U_z\quad\text{for some }\nu\ge1.
\end{equation}
In particular, as $z\in\overline{B_a(x,\delta)}\subset C'$, $w_z\in W_C$.
\end{lemma}

\begin{proof}
In normal coordinates centered at $x$, write $q=\exp_x\xi$.
Inequality \eqref{eq:semiconc} and \eqref{eq:superdiff} give, uniformly,
\[
 b(q)\le b(x)-g_x(v,\xi)+A\norma{\xi}^2,
 \qquad v\in\co U_x.
\]
For past causal $\xi$, compactness of the future timelike family gives
$-g_x(v,\xi)\le-c\norma{\xi}$ with $c>0$ independent of $x$ and $v$.
After shrinking $\delta$ the right-hand side is at most
$b(x)-(c/2)\norma{\xi}$.

For any $u\in U_x$, the point $\exp_x(-ru)$ is an admissible competitor
for small $r$ and has $b$-value at least $b(x)-Lr$, with uniform $L$.
The admissible set is nonempty and compact, so a maximum exists, and every
maximizer satisfies $\norma{\xi}\le2Lr/c$. For small $r$ it lies in the
interior of the auxiliary ball. Uniform bounds for $\exp_x$ also give
the auxiliary length estimate for the radial geodesic.

If $d(z,x)>r$, moving a little from $z$ toward $x$ along that geodesic
preserves admissibility and strictly increases $b$ by
\eqref{eq:rushing}. Thus $d(z,x)=r$.

For the multiplier, let $\psi=d(\cdot,x)$ near $z$; then
$\dd\psi_z=g_z(w_z,\cdot)$. If $D^+b(z)$ were disjoint from the negative
normal ray $\{-\nu\dd\psi_z:\nu\ge0\}$, strong separation would give
$Y\in T_zM$ with $P(Y)>0$ for all $P\in D^+b(z)$ and
$\dd\psi_z(Y)\ge0$. Perturb $Y$ by a sufficiently small multiple of
$-w_z$ to make the latter inequality strict. Upper semicontinuity of
$D^+b$ gives a common positive lower bound for $P(Y)$ at nearby points.
The coordinate segment $z+tY$ is admissible for small $t>0$. At almost
every parameter where its Lipschitz $b$-restriction is differentiable,
\eqref{eq:semiconc} with negative increments shows that its derivative is
at least that lower bound. Hence $b$ increases, contradicting maximality.
Therefore $-\nu\dd\psi_z\in D^+b(z)$, and \eqref{eq:superdiff} gives
\eqref{eq:multiplier}. Its Lorentzian norm is $\nu\ge1$.
\end{proof}

\begin{proposition}[Multiplicative persistence along sphere steps]\label{prop:persistence}
For every compact $C\subset O$ there are $\bar r_C,E_C>0$ such that every
sphere step $x\mapsto z$ of Lemma~\ref{lem:step}, with $x\in C$ and
$0<r\le\bar r_C$, satisfies
\begin{equation}\label{eq:persistence}
 \vartheta(z)-1\ge(1-E_Cr)\bigl(\vartheta(x)-1\bigr),
 \qquad E_C\bar r_C\le\tfrac12.
\end{equation}
In particular every step from a singular point ends at a singular point.
\end{proposition}

\begin{proof}
Let $w$ be the future unit tangent at $x$ of the segment from $z$ to $x$,
so that $z=\exp_x(-rw)$ and $x=\exp_z(rw_z)$. By \eqref{eq:multiplier},
$w_z\in W_C$, and parallel transport along segments of auxiliary length
$O(r)$ puts $w$ in a compact family as well, so Lemma~\ref{lem:gauss}
applies uniformly. Choose convex representations
\[
 V(x)=\sum_i\lambda_i u_i,\qquad \nu w_z=\sum_j\mu_j u'_j,
\]
with $u_i\in U_x$, $u'_j\in U_z$, and put $c_i=c(u_i,w)$,
$c'_j=c(u'_j,w_z)$,
\[
 \Theta_x=\sum_i\lambda_ic_i\ge\vartheta(x),\qquad
 \Theta_z=\sum_j\mu_jc'_j=\nu\le\vartheta(z),
\]
by the reverse Cauchy--Schwarz inequality. The support inequalities
\eqref{eq:support} at $x$ and $z$,
\[
 b(z)\le b(x)+\sigma-d(z,\exp_x(\sigma u_i)),\qquad
 b(x)\le b(z)+\sigma-d(x,\exp_z(\sigma u'_j)),
\]
hold for small $r$. Averaging them with the indicated weights, adding, and
applying \eqref{eq:taylor} with the lower sign at $x$ and the upper sign at
$z$, we obtain
\[
 \Theta_x-\Theta_z\le Kr\Bigl\{
  \sum_i\lambda_i(c_i^2-1)+\sum_j\mu_j((c'_j)^2-1)\Bigr\}.
\]
With a uniform bound $c_*$ for the $c_i,c'_j$, $c^2-1\le(c_*+1)(c-1)$
gives, for $a=\Theta_x-1$ and $\beta=\Theta_z-1$,
$a-\beta\le er(a+\beta)$ with $e=K(c_*+1)$. Hence, for $er<1$,
$\beta\ge a(1-er)/(1+er)\ge(1-2er)a$, which proves \eqref{eq:persistence}
with $E_C=2e$, after shrinking the radius.
\end{proof}

\begin{proposition}[Sphere chains]\label{prop:chains}
For every $p\in\cS$, chains of sphere steps of length at most $h$
starting at $p$, constructed below with arbitrary choices of the maximizing
points, converge locally uniformly as $h\to0$, parametrized by $T$, to the
curve $\gamma(\tau)=\Psi_{T(p)-\tau}(p)$, which has all the properties
stated in Theorem~\ref{thm:main}.
\end{proposition}

\begin{proof}
For integers $m\ge1$ put
\[
 C_m=J^-(p)\cap T^{-1}([T(p)-m,T(p)]),\qquad
 \ell_m=\max_{q\in C_m}d(q,p),
\]
which are compact subsets of $O$. Choose step radii $r_m>0$ and persistence
constants $E_m$ using the preceding results on $C_{m+1}$, with $r_m$
nonincreasing, $E_m$ nondecreasing, $E_mr_m\le1/2$, and $r_m$ so small that
a step from $C_m$ has temporal drop at most one, so that its endpoints lie
in $C_{m+1}$.

{\em Discrete chains and their strength.} Start with $z_0=p$. If $z_k$ has
been defined, let $m(k)$ be the least integer with $z_k\in C_{m(k)}$, set
$s_k=\min\{h,r_{m(k)}\}$, and let $z_{k+1}\ll z_k$ be any maximizing sphere
step of Lorentzian length $s_k$. Before leaving $C_m$, $s_k\ge\min\{h,r_m\}$,
and the unit tangents of the segments belong to a compact family, so their
temporal drops are at least $c_ms_k$ with $c_m>0$. Hence the chain leaves
each $C_m$ after finitely many steps, and its $T$-coordinates tend to
$-\infty$. If $z_k\in C_m$, all preceding nodes belong to $C_m$, and
$\sum_{j<k}s_j\le d(z_k,p)\le\ell_m$. Using $1-a\ge e^{-2a}$ for
$0\le a\le1/2$ in \eqref{eq:persistence},
\begin{equation}\label{eq:kappa}
 \vartheta(z_k)-1\ge
 e^{-2E_m\ell_m}\bigl(\vartheta(p)-1\bigr)=:\kappa_m>0,
\end{equation}
a bound independent of $h$.

{\em Compactness and singularity of the limit.} Join successive nodes by
their timelike geodesic segments and parameterize the resulting broken
curve $\beta_h$ by $T$, in the future direction. On $[T(p)-m,T(p)]$ its
image lies in $C_m$, and its $T$-normalized tangents are uniformly bounded
and separated from the light cone, by \eqref{eq:multiplier}, compactness of
$W_C$ and parallel transport along segments of auxiliary length $O(h)$. The
Arzel\`a--Ascoli theorem and a diagonal subsequence give locally uniform
convergence to a locally Lipschitz curve $\gamma$ with
$T(\gamma(\tau))=\tau$ and $\gamma(T(p))=p$. For
$\tau\in[T(p)-m,T(p)]$, the future endpoint of the segment containing
$\beta_h(\tau)$ is a node in $C_m$ at auxiliary distance $O(h)$; these nodes
converge to $\gamma(\tau)$ and satisfy \eqref{eq:kappa}, so upper
semicontinuity of $\vartheta$ gives $\vartheta(\gamma(\tau))\ge1+\kappa_m$.
No singularity is asserted for the interiors of the approximating segments.

{\em The limiting inclusion.} At the past endpoint $z$ of each step the
future $T$-normalized tangent belongs to $\cF_z$, by \eqref{eq:multiplier},
and in local coordinates the normalized tangent at any point of the step
differs from it by $O(h)$, uniformly on compact temporal slabs; the past
endpoint is also at distance $O(h)$ from the evaluation point. Fix a
differentiability parameter $\tau_0$ of $\gamma$ and a chart around
$x_0=\gamma(\tau_0)$. By upper semicontinuity of $\cF$, for every
$\epsilon>0$ all nearby endpoint velocity sets lie in
$\cF_{x_0}+\epsilon\overline B$. For small intervals around $\tau_0$ and
small $h$, the velocities of $\beta_h$ lie almost everywhere in the closed
convex set $\cF_{x_0}+2\epsilon\overline B$, and so do the coordinate
difference quotients. Passing to the limit, shrinking the interval and
letting $\epsilon\downarrow0$ gives $\dot\gamma(\tau_0)\in\cF_{x_0}$. With
Lemma~\ref{lem:F} and $T(\gamma(\tau))=\tau\to-\infty$, the limit has the
properties stated in Theorem~\ref{thm:main}.

{\em Convergence of the whole family.} Every limit obtained in this way is
a past-directed $C^b$-causal curve from $p$ parametrized by $T$, hence it
is $\tau\mapsto\Psi_{T(p)-\tau}(p)$ by Theorem~\ref{thm:unique}(1).
Since every sequence $h_j\to0$ has a subsequence along which the chains
converge, and all these limits coincide, the whole family converges as
$h\to0$, whatever the choices of the maximizing points.
\end{proof}

\begin{remark}
The limit step is an instance of the closure of causal curves under limits
\cite[Theorem~2.12]{minguzzi17}, applied to the decreasing closed cone
structures generated, in local trivializations, by the convex hulls of the
sets $\cF_{q'}$, $d_a(q',q)\le1/k$, whose intersection is $C^b$. The curve
$\gamma$ is $C^b$-causal, so Theorem~\ref{thm:invariance} and
Lemma~\ref{lem:rigidity} apply to it as well. The construction of the
chains and the compactness and limiting-inclusion arguments do not use
$p\in\cS$, which enters only through \eqref{eq:kappa}; thus, for every
$p\in O$, sphere chains from $p$ converge to the trajectory
$s\mapsto\Psi_s(p)$.
\end{remark}

\section{Calibrated rays, initial points and calibrated lines}\label{sec:rays}

\subsection{The regular phase and maximal calibrated rays}

Recall that a causal curve $x$ in $O$ is $b$-calibrated if
$b(x(t_2))=b(x(t_1))+\ell(x|_{[t_1,t_2]})$ for $t_1<t_2$, where $\ell$ is
the Lorentzian length \cite[Definition~5.2]{minguzzi26e}. Every future
inextendible $b$-calibrated causal curve is a timelike ray, hence, in
proper time, a future complete timelike geodesic
\cite[Theorem~5.12(i)]{minguzzi26e}; these are the calibrated rays of
fact~2. Moreover, $b$ is differentiable at every point $\gamma(t)$, $t>0$,
of every calibrated ray $\gamma$ \cite[Proposition~5.19]{minguzzi26e}. For
$p\in O$ put
\[
 s_0(p)=\inf\{s\ge0:\Psi_s(p)\in\cS\}\in[0,\infty],
 \qquad \inf\varnothing=\infty .
\]

\begin{theorem}[Calibrated rays and the regular phase]\label{thm:regular}
Let $p\in O$.
\begin{enumerate}
\item $\Psi_s(p)\notin\cS$ for $0\le s<s_0(p)$, and $\Psi_s(p)\in\cS$ for
      $s>s_0(p)$.
\item If $\gamma$ is a calibrated ray starting at $q\in O$, then
      $\Psi_{T(\gamma(t))-T(q)}(\gamma(t))=q$ for every $t\ge0$: the
      trajectory of $\Psi$ from $\gamma(t)$ runs backward along $\gamma$
      until $q$.
\item For every finite $s$ with $0\le s\le s_0(p)$, a calibrated ray
      starting at $\Psi_s(p)$ passes through $p$, and its segment from
      $\Psi_s(p)$ to $p$ is the trajectory of $\Psi$ from $p$, reversed.
      Consequently
      \begin{equation}\label{eq:s0}
       s_0(p)=\sup\{T(p)-T(q):\text{a calibrated ray starting at $q$
       passes through $p$}\},
      \end{equation}
      and if $s_0(p)<\infty$, then $\Psi_{s_0(p)}(p)$ belongs to the
      closure of $\cS$ in $O$.
\item If $s_0(p)=\infty$, the trajectory of $\Psi$ from $p$, reversed and
      followed by the calibrated ray starting at $p$, is a timelike line.
      This line is complete if $(M,g)$ is past timelike complete.
\end{enumerate}
\end{theorem}

\begin{proof}
(1) The first assertion is the definition of $s_0(p)$. If $s>s_0(p)$,
there is $s_1\in[s_0(p),s)$ with $\Psi_{s_1}(p)\in\cS$, and
Theorem~\ref{thm:invariance}, applied to the trajectory in proper time,
gives $\Psi_s(p)=\Psi_{s-s_1}(\Psi_{s_1}(p))\in\cS$.

(2) Parametrize $\gamma$ by proper time. For $t>0$, $b$ is differentiable
at $\gamma(t)$, so that $U_{\gamma(t)}$ is a singleton by
Lemma~\ref{lem:strength}; since $\gamma|_{[t,\infty)}$ is a calibrated ray
starting at $\gamma(t)$, $U_{\gamma(t)}=\{\dot\gamma(t)\}$ and
$C^b_{\gamma(t)}=\R_{>0}\dot\gamma(t)$. Hence, for $t_1>0$, the curve
$r\mapsto\gamma(t_1-r)$, $r\in[0,t_1]$, has velocity in
$-C^b_{\gamma(t_1-r)}$ for every $r<t_1$: it is a past-directed
$C^b$-causal curve from $\gamma(t_1)$ to $q$. By
Theorem~\ref{thm:unique}(1), it is the trajectory of $\Psi$ from
$\gamma(t_1)$ up to the drop $T(\gamma(t_1))-T(q)$ of $T$.

(3) Let $x$ be the proper-time parametrization of $u\mapsto\Psi_u(p)$, and
let $x(r_s)=\Psi_s(p)$. Since $\Psi_u(p)\notin\cS$ for $u<s_0(p)$, we have
$\vartheta(x(a))=1$ for every $a\in[0,r_s)$, and \eqref{eq:defect} gives
$b(p)-b(x(r))=r$ for $r\in[0,r_s]$. Let $\gamma$ be a calibrated ray
starting at $p$, parametrized by proper time, and let $c$ be the curve
which follows $r\mapsto x(r_s-r)$, $r\in[0,r_s]$, and then $\gamma$. The
increase of $b$ along every segment of $c$ equals its Lorentzian length, so
$c$ is a future inextendible $b$-calibrated causal curve starting at
$\Psi_s(p)$, hence a calibrated ray \cite[Theorem~5.12(i)]{minguzzi26e}. This
proves the first statement. If $s_0(p)<\infty$, it gives, with
$s=s_0(p)$, a calibrated ray from a point $q$ with $T(p)-T(q)=s_0(p)$; if
$s_0(p)=\infty$, it gives such rays with $T(p)-T(q)$ arbitrarily large.
Conversely, if a calibrated ray $\gamma$ starting at $q$ passes through
$p=\gamma(t_1)$ with $t_1>0$, then by (2) the points $\Psi_u(p)$,
$0\le u<T(p)-T(q)$, are points $\gamma(t)$ with $t>0$, hence
differentiability points of $b$, so that $s_0(p)\ge T(p)-T(q)$. This
proves \eqref{eq:s0}. Finally, if $s_0(p)<\infty$, then $\Psi_{s_0(p)}(p)$
is the limit of the points $\Psi_s(p)\in\cS$ as $s\downarrow s_0(p)$, by
(1) and continuity.

(4) By (3), for every $s\ge0$ the reversed trajectory from $\Psi_s(p)$ to
$p$, followed by the calibrated ray starting at $p$ (which is unique, as
$p\notin\cS$), is a calibrated ray, hence a timelike geodesic along which
$b$ increases by the Lorentzian length. Thus the whole curve is a timelike
geodesic, and by the rushing inequality \eqref{eq:rushing} the Lorentzian
distance between any two of its points is at most the increase of $b$,
which equals the length of the segment between them; as the length never
exceeds the distance, every segment is maximizing. It is future inextendible, as the calibrated ray is, and past
inextendible, as the trajectory of $\Psi$ is (proof of
Theorem~\ref{thm:unique}(2)). Hence it is a timelike line. Its future half
is future complete by the standing assumption, and its past half is a past
inextendible timelike geodesic, which is past complete if $(M,g)$ is past
timelike complete.
\end{proof}

Before $s_0(p)$ the semi-flow therefore moves backward along calibrated
rays, with $\vartheta=1$; after $s_0(p)$ it moves inside $\cS$, with
$\vartheta>1$. If $s_0(p)<\infty$, the point $\Psi_{s_0(p)}(p)$ belongs
to the closure of $\cS$, and it may be regular
(Remark~\ref{rem:threshold}).

\begin{proposition}[The entry time]\label{prop:entry}
For $s>0$ and $p\in O$ put
\[
 D_s(p)=b(p)-b(\Psi_s(p))-d(\Psi_s(p),p).
\]
Then $D_s$ is continuous and nonnegative on $O$, and
\begin{equation}\label{eq:entry}
 \{p\in O:s_0(p)\ge s\}=\{p\in O:D_s(p)=0\}.
\end{equation}
Hence $s_0:O\to[0,\infty]$ is upper semicontinuous. Moreover,
$s_0(\Psi_s(p))=\max\{s_0(p)-s,0\}$ for $s\ge0$, with $\infty-s=\infty$.
\end{proposition}

\begin{proof}
The function $D_s$ is continuous, as $b$ and $\Psi$ are and the Lorentzian
distance is continuous on globally hyperbolic spacetimes, and it is
nonnegative by the rushing inequality \eqref{eq:rushing}, as
$\Psi_s(p)\le p$. If $s_0(p)\ge s$, by Theorem~\ref{thm:regular}(3) the
trajectory from $p$ to $q=\Psi_s(p)$, reversed, is a segment of a
calibrated ray, so that $b(p)-b(q)=d(q,p)$ and $D_s(p)=0$. Conversely, let
$D_s(p)=0$. Since $q\ll p$, there is a maximizing timelike geodesic
$\sigma$ from $q$ to $p$, and $b(p)-b(q)=\ell(\sigma)$. By the rushing
inequality the increase of $b$ along each subsegment of $\sigma$ is at
least its length, and these increases add up to $\ell(\sigma)$, so that
$\sigma$ is $b$-calibrated. Followed by a calibrated ray starting at $p$,
it gives a future inextendible $b$-calibrated curve starting at $q$, hence
a calibrated ray \cite[Theorem~5.12(i)]{minguzzi26e}. By
Theorem~\ref{thm:regular}(2) the trajectory of $\Psi$ from $p$ runs
backward along this ray until $q$, through points which are not the
initial point of the ray, hence through differentiability points of $b$
\cite[Proposition~5.19]{minguzzi26e}. Thus $\Psi_u(p)\notin\cS$ for $u<s$, and
$s_0(p)\ge s$. The sets \eqref{eq:entry} are therefore closed for every
$s>0$, which is the upper semicontinuity of $s_0$. The last identity
follows from the semigroup law and Theorem~\ref{thm:regular}(1).
\end{proof}

\begin{remark}[A regular entry point]\label{rem:threshold}
The point $\Psi_{s_0(p)}(p)$ can be regular, as the following local model
shows. In $1+1$ Minkowski space, $g=-\dd t^2+\dd z^2$, let $a>0$ and
\[
 b(t,z)=\min_{p\in[-a,a]}\bigl\{t\sqrt{1+p^2}-pz+p^4\bigr\},
\]
a minimum of a compact family of affine functions with past unit timelike
gradients; the active parameters $p$ correspond to the vectors
$u_p=(\sqrt{1+p^2},p)$, and Lemma~\ref{lem:danskin} gives the
superdifferential. On the axis $z=0$ one minimizes the even function
$f_t(p)=t\sqrt{1+p^2}+p^4$. For $t\ge0$ its only minimizer is $p=0$, so
that $(t,0)$ is regular. For $t<0$ close to $0$,
$f_t'(p)=p\bigl(t/\sqrt{1+p^2}+4p^2\bigr)$, and the minimizers are exactly
$\pm p(t)$, where $4p(t)^2\sqrt{1+p(t)^2}=-t$. The segment joining
$u_{\pm p(t)}$ has maximal-norm element $V=(\sqrt{1+p(t)^2},0)$, so that
$(t,0)$ is singular and the past-directed $C^b$-causal curve normalized by
$T=t$ moves along the axis. The trajectory starting at $(t_0,0)$, $t_0>0$,
thus has $s_0=t_0$, and its entry point $(0,0)$ is regular and belongs to
$\overline\cS$. This is a local model, not an asserted Cauchy Busemann
function of a spacetime with compact Cauchy hypersurfaces.
\end{remark}

\begin{corollary}[Calibrated lines and maximal calibrated rays]\label{cor:structure}
Let $\mathcal L_b=\{p\in O:s_0(p)=\infty\}$ and
$\mathcal E=\{p\in O:s_0(p)=0\}$, and call a calibrated ray maximal if it
is not a proper final portion of another calibrated ray.
\begin{enumerate}
\item If $\gamma$ is a calibrated ray starting at $q$, then
      $s_0(\gamma(t))=s_0(q)+T(\gamma(t))-T(q)$ for $t\ge0$, with
      $\infty+r=\infty$.
\item A point $p\in O$ belongs to $\mathcal L_b$ if and only if it lies on
      a $b$-calibrated timelike line. Through each point of $\mathcal L_b$
      passes exactly one such line, $\mathcal L_b$ consists of regular
      points, and
      $\mathcal L_b=\bigcap_{m\ge1}\{p\in O:D_m(p)=0\}$ is closed in $O$.
\item $\mathcal E$ is the set of the points of $O$ which are not interior
      points of a calibrated ray, and $\cS\subset\mathcal E\subset\overline\cS$,
      the closure being taken in $O$. A calibrated ray is maximal if and
      only if its initial point belongs to $\mathcal E$.
\item Every $p\in O\setminus(\mathcal L_b\cup\mathcal E)$ is an interior
      point of exactly one maximal calibrated ray, namely the calibrated ray
      starting at $\Psi_{s_0(p)}(p)\in\mathcal E$ and passing through $p$,
      and every calibrated ray through $p$ is a final portion of it.
\end{enumerate}
\end{corollary}

\begin{proof}
(1) Let $\sigma=T(\gamma(t))-T(q)$. By Theorem~\ref{thm:regular}(2),
$\Psi_\sigma(\gamma(t))=q$, and the points $\Psi_u(\gamma(t))$,
$0\le u<\sigma$, are points $\gamma(t')$ with $t'>0$, hence differentiability
points of $b$ \cite[Proposition~5.19]{minguzzi26e}; thus
$s_0(\gamma(t))\ge\sigma$. The identity of Proposition~\ref{prop:entry}
gives $s_0(q)=\max\{s_0(\gamma(t))-\sigma,0\}=s_0(\gamma(t))-\sigma$.

(2) If $p$ lies on a $b$-calibrated timelike line $\ell$, then for every
point $q$ of $\ell$ before $p$ the part of $\ell$ after $q$ is a calibrated
ray starting at $q$ and passing through $p$, and $T(q)\to-\infty$ as $q$
moves to the past along $\ell$, which is past inextendible; hence
$s_0(p)=\infty$ by \eqref{eq:s0}. The converse is
Theorem~\ref{thm:regular}(4), whose line is $b$-calibrated. If $\ell$ is a
$b$-calibrated timelike line through $p\in\mathcal L_b$, its part after $p$
is the calibrated ray starting at $p$, which is unique as $p\notin\cS$, and
by Theorem~\ref{thm:regular}(2) each point $q$ of $\ell$ before $p$ equals
$\Psi_{T(p)-T(q)}(p)$; thus $\ell$ is determined by $p$. Every
$p\in\mathcal L_b$ is regular by Theorem~\ref{thm:regular}(1), and
$\mathcal L_b=\bigcap_m\{s_0\ge m\}$ is closed by \eqref{eq:entry}.

(3) If a calibrated ray starting at $q\ne p$ passes through $p$, then
$q\ll p$, so that $T(q)<T(p)$; thus, by \eqref{eq:s0}, $s_0(p)=0$ if and
only if $p$ is not an interior point of a calibrated ray. Hence
$\cS\subset\mathcal E$ by \cite[Proposition~5.19]{minguzzi26e}. If
$p\in\mathcal E$, then $\Psi_s(p)\in\cS$ for $s>0$, by
Theorem~\ref{thm:regular}(1), and $\Psi_s(p)\to p$, so that
$p\in\overline\cS$. A calibrated ray starting at $q\in\mathcal E$ is not a
proper final portion of another calibrated ray, since $q$ would be an
interior point of the latter. If $q\notin\mathcal E$, a calibrated ray
starting at some $q'\ne q$ passes through $q$; as $q\notin\cS$, the
calibrated ray starting at $q$ is unique, and it is the final portion of
that ray after $q$.

(4) Let $q=\Psi_{s_0(p)}(p)$, so that $0<T(p)-T(q)=s_0(p)<\infty$. By
Proposition~\ref{prop:entry}, $s_0(q)=0$, that is, $q\in\mathcal E$, and by
Theorem~\ref{thm:regular}(3) a calibrated ray $\gamma_0$ starting at $q$
passes through $p$, which is an interior point of it; $\gamma_0$ is maximal
by (3). Let $\gamma$ be a calibrated ray through $p$, starting at $q'$. By
Theorem~\ref{thm:regular}(2), $q'=\Psi_{T(p)-T(q')}(p)$, and
$T(p)-T(q')\le s_0(p)$ by \eqref{eq:s0}, so that, by
Theorem~\ref{thm:regular}(3), $q'$ lies on the segment of $\gamma_0$ from $q$
to $p$. If $q'\ne q$, then $q'$ is an interior point of $\gamma_0$, hence
regular, and $\gamma$ is the final portion of $\gamma_0$ after $q'$. If
$q'=q$, then $p$ is an interior point of both $\gamma$ and $\gamma_0$, whose
parts after $p$ coincide by \cite[Proposition~5.6]{minguzzi26e}; being
geodesics, $\gamma$ and $\gamma_0$ coincide. Hence every calibrated ray
through $p$ is a final portion of $\gamma_0$, which is the only maximal one.
\end{proof}

Thus $O$ is the disjoint union of $\mathcal L_b$, a union of $b$-calibrated
timelike lines, of the set $\cE$ of initial points, which lies between $\cS$
and its closure, and of the interiors of the maximal calibrated rays, which
are pairwise disjoint.

\begin{remark}[The canonical calibrating semi-flow]\label{rem:canonical}
Let $b=b^{\textnormal{\textsc{m}}}$, as allowed in Section~\ref{sec:setting}, and let
$\Phi$ and $D$ be the canonical calibrating semi-flow and the set of
differentiability points of $b^{\textnormal{\textsc{m}}}$ of
\cite[Section~5.1]{minguzzi26e}, so that $\Phi_s(p)$ is the point at proper
time $s$ of the unique calibrated ray starting at $p\in D$. By
Theorem~\ref{thm:regular}(2),
\[
 \Psi_{T(\Phi_s(p))-T(p)}\bigl(\Phi_s(p)\bigr)=p
 \qquad(p\in D,\ s\ge0):
\]
run for the corresponding drop of $T$, the past semi-flow inverts $\Phi_s$;
in proper time, the trajectory of $\Psi$ from $\Phi_s(p)$ reaches $p$ after
proper time $s$. Unlike $\Phi$, the semi-flow $\Psi$ is defined on the whole
of $\dom b^{\textnormal{\textsc{m}}}$, and it also undoes the branching of
the calibrated rays at singular points: the points of all the calibrated
rays starting at $q\in\cS$ are mapped back to $q$. Moreover,
$\mathcal L_{b^{\textnormal{\textsc{m}}}}\ne\varnothing$
(Corollary~\ref{cor:depth}(2)), so that $\Phi$ has orbits which extend to
calibrated timelike lines emanating from the initial singularity.
\end{remark}

\subsection{Initial points and lower supports}

\begin{definition}
Let $p\in O$. A {\em smooth lower support} of $b$ at $p$ is a smooth
function $\phi$, defined on a neighbourhood of $p$, with $\phi(p)=b(p)$ and
$\phi\le b$. A covector $P\in T^*_pM$ is a {\em proximal subgradient} of $b$
at $p$ if, in a chart around $p$, there is $C\ge0$ such that
\begin{equation}\label{eq:prox}
 b(y)\ge b(p)+P(y-p)-C|y-p|^2
\end{equation}
for $y$ near $p$. The set $D^-_{\mathrm{prox}}b(p)$ of proximal subgradients
is nonempty if and only if $b$ admits a smooth lower support at $p$, a
property which does not depend on the chart. For the restriction of $b$ to
a $C^2$ hypersurface $\Sigma'$ through $p$ we use quadratic lower supports
in $C^2$ coordinates on $\Sigma'$.
\end{definition}

\begin{lemma}[A small spacelike cap]\label{lem:cap}
Let $\Sigma'$ be a $C^2$ spacelike hypersurface through $p\in M$, let $u$ be
a future unit timelike vector at $p$, let $\xi\mapsto z(\xi)\in\Sigma'$ be
$C^2$ coordinates with $z(0)=p$, and put $P=-g_p(u,\dd z_0(\cdot))$ and
$q_r=\exp_p(-ru)$. After shrinking the coordinate domain to a ball $B_R$,
there are $r_0,c,C>0$ such that, for $0<r\le r_0$, the set $D_r$ of the
$\xi\in B_R$ with $z(\xi)$ in the timelike future of $q_r$ within a fixed
causally convex, convex normal neighbourhood of $p$ is convex, contains $0$,
and satisfies
\begin{equation}\label{eq:capbound}
 D_r\subset\{|\xi|\le Cr\},\qquad
 d(q_r,z(\xi))\le r+P\xi-\frac cr|\xi|^2\quad(\xi\in\overline{D_r}).
\end{equation}
If moreover $p\in O$ and $u\in U_p$, then, for $r$ small, every calibrated
ray starting at $q_r$ meets $z(D_r)$. The constants $R,r_0,c,C$, and the
smallness of $r$ in the last statement, can be chosen uniformly for $p$ in
a compact set, $u$ in a compact set of future unit timelike vectors at the
points $p$, and parametrizations $z$ ranging in a compact set in the $C^2$
topology.
\end{lemma}

\begin{proof}
Let $N$ be a causally convex, convex normal neighbourhood of $p$. The
function $W(q,z)=-g_q(\exp_q^{-1}z,\exp_q^{-1}z)$ is smooth on $N\times N$,
and $W(q,z)=d(q,z)^2$ if $q\ll z$, since local and global distance agree on
the causally convex set $N$. Put $W_r(\xi)=W(q_r,z(\xi))$, a $C^2$ function
of $\xi$ which depends continuously on $r$ in the $C^2$ topology. The first
variation formula gives $W_r(0)=r^2$ and $DW_r(0)=2rP$. At $r=0$,
$\exp_p^{-1}z(\xi)=\dd z_0\xi+O(|\xi|^2)$, so that $D^2W_0(0)=-2z^*g_p$,
which is negative definite as $\Sigma'$ is spacelike. Hence, after
shrinking to a ball $B_R$, there are $\kappa,r_0>0$ with
$D^2W_r\le-\kappa I$ on $B_R$ for $0\le r\le r_0$, and
\[
 W_r(\xi)\le r^2+2rP\xi-\frac\kappa2|\xi|^2 .
\]
The set $\{\xi\in B_R:W_r(\xi)>0\}$ is convex, as a superlevel set of a
concave function, contains $0$, and by the last inequality is contained in
$\{|\xi|\le Cr\}$, on which $W_r\le M^2r^2$, for constants $C,M$; after
decreasing $r_0$, $\{|\xi|\le Cr\}\subset B_R$. On this connected set the
timelike vector $\exp^{-1}_{q_r}z(\xi)$ keeps its time orientation, which
is future at $\xi=0$, so the set is $D_r$, and $d(q_r,z(\xi))=d_r(\xi):=
W_r(\xi)^{1/2}$ there. Moreover
\[
 D^2d_r=\frac{D^2W_r}{2d_r}-\frac{DW_r\otimes DW_r}{4d_r^3}
 \le-\frac{\kappa}{2Mr}I ,
\]
and since $d_r(0)=r$ and $Dd_r(0)=P$, Taylor's formula along the segment
from $0$ to $\xi$ gives \eqref{eq:capbound} on $D_r$, with $c=\kappa/(4M)$.
At the points of $\partial D_r$ the distance vanishes, and continuity of $d$
extends the inequality to $\overline{D_r}$.

For the last statement take coordinates $(\tau,\xi)$ near $p$ with
$\Sigma'=\{\tau=0\}$ and $\dd\tau$ positive on future causal vectors, and a
compact neighbourhood $K\subset N\cap O$ of $p$. The unit tangents of
calibrated rays at points of $K$ form a compact family (fact~2), so their
coordinate norms are bounded and $\dd\tau\ge m>0$ on them. As
$\tau(q_r)=-r\,\dd\tau_p(u)+O(r^2)$, a calibrated ray starting at $q_r$
reaches $\{\tau=0\}$ within proper time $O(r)$, before it can leave $K$,
at a point at coordinate distance $O(r)$ from $p$. For small $r$ this point
belongs to $\Sigma'$ and to the timelike future of $q_r$ within $N$, hence
to $z(D_r)$. All the constants come from continuity and compactness, which
gives the uniformity statement.
\end{proof}

\begin{theorem}[Initial points and lower supports]\label{thm:support}
For $p\in O$ the following conditions are equivalent:
\begin{enumerate}
\item $p\notin\cE$, that is, $p$ is an interior point of a calibrated ray;
\item $b$ admits a smooth lower support at $p$;
\item $D^-_{\mathrm{prox}}b(p)\ne\varnothing$;
\item for some, equivalently for every, $C^2$ spacelike hypersurface
      $\Sigma'$ through $p$, the restriction $b|_{\Sigma'}$ admits a
      quadratic lower support at $p$.
\end{enumerate}
In particular, $\cE=\{p\in O:D^-_{\mathrm{prox}}b(p)=\varnothing\}$.
\end{theorem}

\begin{proof}
(1)$\Rightarrow$(2). By Corollary~\ref{cor:structure}(3), $p$ is an interior
point of a calibrated ray $\gamma$. Let $q$ be a point of $\gamma$ before
$p$, so close to $p$ that both lie in a causally convex, convex normal
neighbourhood. Then $\phi=b(q)+d(q,\cdot)$ is smooth near $p$, $\phi\le b$
there by the rushing inequality \eqref{eq:rushing}, and $\phi(p)=b(p)$
because $\gamma$ is calibrated. The equivalence (2)$\Leftrightarrow$(3) was
noted in the definition, and restricting a smooth lower support gives (4)
for every $\Sigma'$.

(4)$\Rightarrow$(1). Let $b(z(\xi))\ge b(p)+P\xi-C_0|\xi|^2$ in $C^2$
coordinates $z$ on $\Sigma'$ with $z(0)=p$. By \eqref{eq:semiconc}, the
restriction $b\circ z$ is semiconcave, with upper supports $P'\xi+A|\xi|^2$
for every $P'$ in its superdifferential at $0$; together with the lower
support, this forces $P'=P$. By \eqref{eq:superdiff}, the covectors
$-g_p(u,\dd z_0(\cdot))$, $u\in U_p$, belong to that superdifferential, so
they all equal $P$. A future unit timelike vector $u$ is determined by the
restriction of $g_p(u,\cdot)$ to $T_p\Sigma'$: writing $u=\alpha\nu+w$, with
$\nu$ the future unit normal and $w\in T_p\Sigma'$, the restriction
determines $w$, and $\alpha=\sqrt{1+g(w,w)}$. Hence $U_p=\{u\}$.

Apply Lemma~\ref{lem:cap} to this $u$, with $r$ so small that
$c/r\ge C_0$. By \eqref{eq:capbound} and the lower support,
\[
 b(z)-d(q_r,z)\ge b(p)-r\qquad(z\in z(D_r)).
\]
A calibrated ray starting at $q_r$ meets $z(D_r)$ at a point $z_r$, and
$b(z_r)-b(q_r)=d(q_r,z_r)$, as calibrated segments are maximizing; thus
$b(q_r)\ge b(p)-r$. On the other hand the rushing inequality gives
$b(q_r)\le b(p)-d(q_r,p)\le b(p)-r$. Hence $b(p)-b(q_r)=r$, the length of
the geodesic segment from $q_r$ to $p$, and, by the rushing inequality on
its subsegments, this segment is $b$-calibrated. Followed by a calibrated
ray starting at $p$, it gives a future inextendible $b$-calibrated curve
starting at $q_r$, hence a calibrated ray \cite[Theorem~5.12(i)]{minguzzi26e},
of which $p$ is an interior point.
\end{proof}

\begin{remark}[The analytic cut locus]
For a locally semiconcave function with linear modulus, Cannarsa, Cheng,
Hong and Wei define the cut locus as the set of points at which the function
cannot be touched from below by a semiconvex function, and they show that it
is the set of points with empty proximal subdifferential
\cite[Definition~3.1, Proposition~3.1 and Remark~3.1]{CCHW}.
Theorem~\ref{thm:support} identifies the geometrically defined set $\cE$
with this analytic cut locus of $b$. Its proof uses Lorentzian distance
comparison, and no Lax--Oleinik representation of $b$.
\end{remark}

\begin{theorem}[Measure of the initial points]\label{thm:measure}
For every $C^2$ spacelike hypersurface $\Sigma'\subset M$,
$\mathcal H^n(\cE\cap\Sigma')=0$. Moreover, $\mathcal H^{n+1}(\cE)=0$.
\end{theorem}

\begin{proof}
We use Alexandrov's theorem on the second order differentiability of convex
functions \cite{alexandrov39}, \cite[Section~6.4]{EG15}: a convex function
$f$ on an open subset of $\R^m$ admits, at almost every point $x$, a second
order Taylor expansion
\[
 f(x+h)=f(x)+Df(x)h+\tfrac12D^2f(x)(h,h)+o(|h|^2)\qquad(h\to0).
\]
Let $\xi\mapsto z(\xi)$ be $C^2$ coordinates on a portion of $\Sigma'\cap O$,
defined on a ball. By \eqref{eq:semiconc}, and since the superdifferentials
of $b$ are locally bounded, there is $A'$ with
$b(z(\eta))\le b(z(\xi))+P'(\eta-\xi)+A'|\eta-\xi|^2$ for a covector $P'$
depending on $\xi$. Hence $\xi\mapsto b(z(\xi))-A'|\xi|^2$ has an affine upper
support at every point, so it is concave on a smaller ball, and by
Alexandrov's theorem $b\circ z$ has a second order expansion, hence a
quadratic lower support, at almost every point. Such points do not belong
to $\cE$, by Theorem~\ref{thm:support}. Coordinate Lebesgue null sets are
$\mathcal H^n$-null for the induced, or any auxiliary, Riemannian metric,
and a countable cover proves the first statement. The second follows in
the same way, applying Alexandrov's theorem to $b-A|x|^2$ in spacetime
charts and using condition~(3) of Theorem~\ref{thm:support}.
\end{proof}

\begin{remark}[Horizon interpretation]
Under the identification $x\mapsto(x,b(x))$ of $O$ with the horizon $H$ of
the product spacetime used in \cite{minguzzi26e}, the calibrated rays
correspond to the null generators of $H$ \cite[Proposition~5.6]{minguzzi26e},
which are future inextendible. Thus $\cE$ corresponds to the set of past
endpoints of generators, $\cS$ to the set of points at which $H$ is not
differentiable, and $\cE\setminus\cS$ to the set of endpoints at which $H$
is differentiable. For future horizons, Chru\'sciel, Fu, Galloway and Howard
proved that the latter set has vanishing Hausdorff measure of the dimension
of the horizon, and that its intersection with every $C^2$ spacelike
hypersurface has vanishing measure of one dimension less
\cite[Theorem~1]{CFGH}. Applied to $H$ and to the hypersurfaces
$\Sigma'\times\R$, and combined with $\mathcal H^n(\cS\cap\Sigma')=0$
\cite[Theorem~5.21(c)]{minguzzi26e}, it gives another proof of
Theorem~\ref{thm:measure}. The proof above only uses lower supports and
Alexandrov's differentiability theorem. Neither statement implies the countable
$n$-rectifiability of $\cE$, or the bound $\dim_{\mathcal H}\cE\le n$.
\end{remark}

\subsection{Topology of the initial points}

\begin{proposition}[Topology of the initial points]\label{prop:topology}
The set $\cE=\bigcap_{m\ge1}\{p\in O:s_0(p)<1/m\}$ is a $G_\delta$ subset
of $O$, its closure in $O$ coincides with that of $\cS$, and
$\Psi_s(\cE)\subset\cS$ for $s>0$. The function $s_0$ is continuous at
every point of $\cE$. If $\cE\ne\varnothing$, then $\cE$ is not achronal.
\end{proposition}

\begin{proof}
Each set $\{s_0<1/m\}$ is open by Proposition~\ref{prop:entry}, and their
intersection is $\{s_0=0\}$. The inclusions $\cS\subset\cE\subset\overline\cS$
of Corollary~\ref{cor:structure}(3) give the equality of closures. If
$p\in\cE$ and $s>0$, then $s>s_0(p)$ and $\Psi_s(p)\in\cS$ by
Theorem~\ref{thm:regular}(1); as the trajectory is timelike,
$\Psi_s(p)\in\cS\cap I^-(p)\subset\cE\cap I^-(p)$. Finally, if $p_j\to p\in\cE$,
upper semicontinuity gives $0\le\liminf s_0(p_j)\le\limsup s_0(p_j)\le
s_0(p)=0$.
\end{proof}

\begin{proposition}[The initial-point map]\label{prop:Q}
On the open set $O\setminus\mathcal L_b$ put $Q(p)=\Psi_{s_0(p)}(p)$. Then
$Q(p)\in\cE$, $Q$ is the identity on $\cE$, and $s_0=T-T\circ Q$. For
$q\in\cE$, $Q^{-1}(q)$ is the union of the calibrated rays starting at $q$.
The map $Q$ is continuous at $p$ if and only if $s_0$ is continuous at $p$;
in particular $Q$ is continuous at every point of $\cE$, and $T\circ Q$ is
lower semicontinuous. If $p_j\to p\notin\mathcal L_b$, the cluster points
of $Q(p_j)$ lie on the segment from $Q(p)$ to $p$ of the trajectory of
$p$.
\end{proposition}

\begin{proof}
The set $\mathcal L_b$ is closed by Corollary~\ref{cor:structure}(2), and
$s_0$ is finite on its complement. By Proposition~\ref{prop:entry},
$s_0(Q(p))=0$, and $Q(p)=p$ if $s_0(p)=0$; the identity $s_0=T-T\circ Q$
is the $T$-normalization of $\Psi$. If $\gamma$ is a calibrated ray starting
at $q\in\cE$, then $s_0(\gamma(t))=T(\gamma(t))-T(q)$ by
Corollary~\ref{cor:structure}(1), and
$\Psi_{T(\gamma(t))-T(q)}(\gamma(t))=q$ by Theorem~\ref{thm:regular}(2), so
that $Q(\gamma(t))=q$. Conversely, if $Q(p)=q$, a calibrated ray starting at
$q$ passes through $p$, by Theorem~\ref{thm:regular}(3) with $s=s_0(p)$. If
$s_0$ is continuous at $p$, so is $Q$, by the joint continuity of $\Psi$;
the converse follows from $s_0=T-T\circ Q$. The remaining continuity
statements follow from Propositions~\ref{prop:entry}
and~\ref{prop:topology}. Finally, if $p_j\to p\notin\mathcal L_b$, upper
semicontinuity bounds $s_0(p_j)$ for large $j$; along a subsequence with
$s_0(p_j)\to\tau$ we have $0\le\tau\le s_0(p)$ and
$Q(p_j)\to\Psi_\tau(p)$.
\end{proof}

\begin{remark}[A discontinuous initial-point map]\label{rem:Qdisc}
In the local model of Remark~\ref{rem:closure}, with $T=t$, the map $Q$ is
discontinuous, $s_0$ is not lower semicontinuous, and $\cE$ is not closed.
Fix $t>0$. The singular points $(t,z_k(t))$ belong to $\cE$, so that
$Q(t,z_k(t))=(t,z_k(t))\to(t,0)$. On the other hand,
$s_0(t,0)=t$ and $Q(t,0)=(0,0)$. Indeed, the segment of the calibrated
ray $\alpha$ from $(0,0)$ to $(t,0)$ gives $s_0(t,0)\ge t$ by
\eqref{eq:s0}. Conversely, for every $\tau<0$ a sufficiently small
$p\in P\setminus\{0\}$ satisfies $\tau(\sqrt{1+p^2}-1)+p^4<0$, so that
$b(\tau,0)<\tau$. If $s_0(t,0)>t$, by Theorem~\ref{thm:regular}(3) the
trajectory of $(t,0)$ would follow, beyond the drop $t$, the reversed
calibrated ray through the regular point $(t,0)$, which is the axis $z=0$,
and calibration would give $b(\tau,0)=b(t,0)-(t-\tau)=\tau$ for some
$\tau<0$, a contradiction. The local arguments of
Theorem~\ref{thm:regular} apply to this model. As with the model itself,
this shows sharpness under the local analytic hypotheses, without asserting
a Cauchy Busemann function of a spacetime with compact Cauchy
hypersurfaces. In particular, the decomposition into maximal calibrated
rays does not by itself provide a continuous retraction onto $\cE$.
\end{remark}

\subsection{Capture by the singular set and calibrated lines}

\begin{proposition}[Capture away from calibrated lines]\label{prop:capture}
For every compact $C\subset O\setminus\mathcal L_b$ there is $S_C<\infty$
such that $\Psi_s(C)\subset\cS$ for $s>S_C$. The points of $O$ whose
trajectories never meet $\cS$ are exactly those of $\mathcal L_b$; their
trajectories remain in $\mathcal L_b$ and consist of regular points.
\end{proposition}

\begin{proof}
The function $s_0$ is finite and upper semicontinuous on $C$, so it attains
a finite maximum $S_C$, and $\Psi_s(p)\in\cS$ for $s>S_C\ge s_0(p)$, by
Theorem~\ref{thm:regular}(1). If $p\in\mathcal L_b$, then
$s_0(\Psi_s(p))=\infty$ for every $s\ge0$, by Proposition~\ref{prop:entry},
so the trajectory remains in $\mathcal L_b$, which consists of regular
points by Corollary~\ref{cor:structure}(2).
\end{proof}

\begin{corollary}[Spacetimes without timelike lines]\label{cor:nolines}
Assume that $(M,g)$ contains no $b$-calibrated timelike line, for instance
no timelike line. Then $\mathcal L_b=\varnothing$, and for every compact
$C\subset O$ there is $S_C<\infty$ such that $\Psi_s(C)\subset\cS$ for
$s>S_C$. Every $p\in O$ lies on a calibrated ray starting at the point
$\Psi_{s_0(p)}(p)\in\cE$, which is not an interior point of any calibrated
ray and lies in the closure of $\cS$ in $O$, and every point of
$O\setminus\cE$ is an interior point of exactly one maximal calibrated ray.
In particular $\cS\ne\varnothing$. This applies to every counterexample to
Bartnik's conjecture: if $(M,g)$ is timelike complete, satisfies TCC and
does not split, then a timelike line would be complete and, by the
Lorentzian splitting theorem \cite[Section~1.4]{minguzzi26e}, would force
$(M,g)$ to split.
\end{corollary}

\begin{proof}
By Corollary~\ref{cor:structure}(2), $\mathcal L_b=\varnothing$. The capture
statement is Proposition~\ref{prop:capture}, and the remaining statements
follow from Theorem~\ref{thm:regular}(3) and Corollary~\ref{cor:structure}.
If $\cS$ were empty, then $\cE=\varnothing$ and every point would be an
interior point of a maximal calibrated ray starting on $\cE$, which is
impossible as $O\ne\varnothing$.
\end{proof}

Corollary~\ref{cor:nolines} strengthens \cite[Remark~5.20]{minguzzi26e}, which
gives, under the same assumptions, a single point of $\dom b^+$ from which
at least two calibrated rays start: in a counterexample, every trajectory
of the past semi-flow is eventually captured by $\cS$, uniformly on compact
sets. We do not call $\cS$ an attractor, since it need not be closed and is
in general not compact.

The existence of calibrated lines can be detected from the past. The
following statement is a form of the one-sided criterion
\cite[Theorem~7.4]{minguzzi26e}; here it follows directly from the upper
semicontinuity of the entry time.

\begin{proposition}[Calibrated lines from a bounded past defect]\label{prop:pastdefect}
Let $\Lambda$ be a compact Cauchy hypersurface, and let $p_n\in O\cap
J^-(\Lambda)$ be a sequence without accumulation points in $M$ such that
\[
 C:=\sup_n\bigl[b(p_n)+d(p_n,\Lambda)\bigr]<\infty .
\]
Then $\mathcal L_b\cap\Lambda$ contains a point $x$ with $b(x)\le C$. In
particular, there is a $b$-calibrated timelike line through $x$.
\end{proposition}

\begin{proof}
Discarding finitely many terms, $p_n\in I^-(\Lambda)$, as $\Lambda$ is
compact. A calibrated ray starting at $p_n$ meets $\Lambda$ at a point
$x_n\ne p_n$, which is an interior point of it. Calibrated segments are
maximizing, so that $b(x_n)=b(p_n)+d(p_n,x_n)\le b(p_n)+d(p_n,\Lambda)\le
C$. By \eqref{eq:s0}, $s_0(x_n)\ge T(x_n)-T(p_n)$. For every $a\in\R$ the
set $J^-(\Lambda)\cap T^{-1}([a,\infty))$ is compact, by global
hyperbolicity, so $T(p_n)\to-\infty$, while $T$ is bounded on $\Lambda$;
hence $s_0(x_n)\to\infty$. Passing to a subsequence, $x_n\to x\in\Lambda$.
Since $b$ is continuous as a map to $(-\infty,+\infty]$
\cite[Theorem~5.12]{minguzzi26e}, $b(x)\le C$, so that $x\in O$, and the
upper semicontinuity of $s_0$ (Proposition~\ref{prop:entry}) gives
$s_0(x)=\infty$, namely $x\in\mathcal L_b$. The line is provided by
Corollary~\ref{cor:structure}(2).
\end{proof}

\begin{corollary}[Bounded past depth]\label{cor:depth}
Let $\Lambda$ be a compact Cauchy hypersurface.
\begin{enumerate}
\item If $d(\cdot,\Lambda)$ is bounded along some past inextendible causal
      curve starting from a point of $O$, then $\mathcal L_b\ne\varnothing$,
      and $\mathcal L_b$ meets every Cauchy hypersurface.
\item If the cosmological time is bounded on $\Lambda$, in particular if it
      is regular, then $\mathcal L_b\ne\varnothing$. This holds for the
      canonical function $b^{\textnormal{\textsc{m}}}$ of every spacetime
      with regular cosmological time.
\item If $(M,g)$ contains no $b$-calibrated timelike line, then for every
      $C\in\R$ the set $\{p\in O\cap J^-(\Lambda):b(p)+d(p,\Lambda)\le C\}$
      is compact, and $d(\cdot,\Lambda)\to+\infty$ along every past
      inextendible causal curve contained in $O$.
\end{enumerate}
\end{corollary}

\begin{proof}
(1) Let $\gamma$ start from $p\in O$. As $J^-(O)\subset O$, $\gamma\subset
O$, and by the rushing inequality \eqref{eq:rushing} $b\le b(p)$ along
$\gamma$. The curve $\gamma$ is eventually contained in $J^-(\Lambda)$, and
its points have no accumulation points in the past direction, by strong
causality, so Proposition~\ref{prop:pastdefect} applies. The lines are
inextendible timelike curves, hence they meet every Cauchy hypersurface.
(2) For $y\in J^-(\Lambda)$ and $r\in\Lambda\cap J^+(y)$,
$d(y,r)\le\tau(r)$ by the definition of the cosmological time $\tau$, so
that $d(\cdot,\Lambda)\le\sup_\Lambda\tau$ on $J^-(\Lambda)$, and (1)
applies, as $O\ne\varnothing$. A regular cosmological time is continuous
\cite{andersson98}, hence bounded on $\Lambda$.
(3) The set is closed in $M$, as $b$ is continuous as a map to
$(-\infty,+\infty]$ and $d(\cdot,\Lambda)$ is continuous; if it were not
compact, it would contain a sequence without accumulation points, and
Proposition~\ref{prop:pastdefect} would provide a $b$-calibrated timelike
line. The last statement follows from (1), as $d(\cdot,\Lambda)$ is
non-decreasing toward the past along causal curves.
\end{proof}

Thus, in every future timelike complete spacetime with compact Cauchy
hypersurfaces and an initial singularity of bounded cosmological time, the
past semi-flow has trajectories which are never captured by $\cS$: they run
backward along calibrated timelike lines, which emanate from the initial
singularity, are maximizing between any two of their events, and are
calibrated for $b$ up to the infinite future. For
$b=b^{\textnormal{\textsc{m}}}$ these are eternal maximally aging observers.
They are past incomplete, as their past halves have proper length bounded
by the cosmological time. In particular Corollary~\ref{cor:nolines} never
applies to such spacetimes, and the capture of
Proposition~\ref{prop:capture} is a phenomenon which coexists with the
calibrated lines. Conversely, in a spacetime without $b$-calibrated lines,
such as a counterexample to Bartnik's conjecture, the future cosmological
development has an infinitely deep past, and $b+d(\cdot,\Lambda)$ diverges
uniformly at past infinity.

\begin{corollary}[A Cauchy hypersurface in the development forces a line]\label{cor:cauchyline}
If $O$ contains a Cauchy hypersurface, then $M$ contains a $b$-calibrated
timelike line. In particular, in a counterexample to Bartnik's conjecture
$\dom b$ contains no Cauchy hypersurface.
\end{corollary}

\begin{proof}
Let $\Lambda\subset O$ be a Cauchy hypersurface. As $\Lambda$ is compact,
every level $T^{-1}(c)$ with $c<\min_\Lambda T$ lies in
$J^-(\Lambda)\subset O$, by fact~1. Fix such a $c$, and suppose that $s_0$
is finite on $T^{-1}(c)$. By Proposition~\ref{prop:entry} it is bounded
there by some $S<\infty$, so that $\Psi_s(T^{-1}(c))\subset\cS$ for $s>S$
(Theorem~\ref{thm:regular}(1)). On the other hand,
$\Psi_s(T^{-1}(c))=T^{-1}(c-s)$: every $y\in T^{-1}(c-s)$ lies in $O$, and
the point at level $c$ of a calibrated ray starting at $y$ is mapped to $y$
by $\Psi_s$ (Theorem~\ref{thm:regular}(2)). Thus $T^{-1}(c-s)\subset\cS$,
which contradicts $\mathcal H^n(\cS\cap T^{-1}(c-s))=0$
(Corollary~\ref{cor:geometry}(1)). Hence $s_0(p)=\infty$ for some
$p\in T^{-1}(c)$, and Theorem~\ref{thm:regular}(4) gives a $b$-calibrated
timelike line. The last statement follows as in
Corollary~\ref{cor:nolines}.
\end{proof}

The line need not be past complete under the standing assumptions. Under
timelike completeness, Corollary~\ref{cor:cauchyline} also follows from
\cite{minguzzi26e}: by the time dual of \cite[Proposition~3.7]{minguzzi26e},
$\dom b^-$ meets every Cauchy hypersurface, so that a Cauchy hypersurface
contained in $\dom b^+$ gives $\dom b^+\cap\dom b^-\ne\varnothing$, and
\cite[Theorem~7.2]{minguzzi26e} provides a complete timelike line.
Corollary~\ref{cor:cauchyline} uses only future timelike completeness, as
does the one-sided criterion \cite[Theorem~7.4]{minguzzi26e}, which gives
another proof: if $\Lambda\subset O$ is a Cauchy hypersurface and $\eta$ is a
past $\Lambda$-ray (time dual of \cite[Theorem~2.3]{minguzzi26e}), defined on $(a,0]$,
$a\in[-\infty,0)$, and parametrized with proper time, so
that $\eta(0)\in\Lambda$, $\eta(s)\le\eta(s')$ for $s\le s'$ and
$d(\eta(s),\Lambda)=-s$, then $\eta\subset J^-(\Lambda)\subset O$ and $b(\eta(s))+d(\eta(s),\Lambda)
=b(\eta(s))-s$ is non-decreasing in $s$, by the rushing inequality, hence bounded
by $b(\eta(0))$ for $s\le 0$, so that
Proposition~\ref{prop:pastdefect} applies.

\section{Fibres of the semi-flow and forward calibrated maps}\label{sec:fibres}

\subsection{Fibres of the semi-flow}

We now describe the fibres of $\Psi_s$. Write $\Lambda_c=T^{-1}(c)$, a
compact Cauchy hypersurface. For $y\in O$ and $c\ge T(y)$, let $\cR_c(y)$
be the set of points $z\in\Lambda_c$ for which there is a future-directed
$C^b$-causal curve in $O$ from $y$ to $z$, namely a locally absolutely
continuous curve with velocity in $C^b$ almost everywhere; thus
$\cR_{T(y)}(y)=\{y\}$. It is the slice at level $c$ of the future reachable
set of $y$ for the cone structure $C^b$. A compact subset $R$ of an
$n$-manifold is {\em cellular} if it is the intersection of a decreasing
sequence of closed $n$-cells, each contained in the interior of the
previous one \cite[p.~386]{Uspenskij}. A cellular set is connected and
cell-like: its inclusion into each of its neighbourhoods is
null-homotopic. This does not assert that the set itself is
contractible. A map between
compact metric spaces is a {\em near-homeomorphism} if it is a uniform limit
of homeomorphisms \cite{Uspenskij}.

\begin{theorem}[Fibres of the semi-flow]\label{thm:fibres}
Let $y\in O$, $c\ge T(y)$ and $s=c-T(y)$.
\begin{enumerate}
\item $\Psi_s^{-1}(y)=\cR_c(y)$. For every $c'\in\R$ and $s'\ge0$,
      $\Psi_{s'}$ maps $\Lambda_{c'}\cap O$ onto $\Lambda_{c'-s'}\cap O$.
\item If $y\notin\cS$, then $\cR_c(y)$ is the single point at level $c$ of
      the calibrated ray starting at $y$, and every future-directed
      $C^b$-causal curve starting at $y$ is, up to parametrization, a
      portion of that ray. If $y\in\cS$ and $s>0$, then $\cR_c(y)$ contains
      the points at level $c$ of all the calibrated rays starting at $y$,
      and at least two points. Consequently, for every $s'>0$,
      \[
       \cS=\{y\in O:\Psi_{s'}^{-1}(y)\ \text{has more than one point}\}.
      \]
      For $a<c$ put $D_a=(\Lambda_a\cap O)\setminus\cS$. Then
      $\Psi_{c-a}$ restricts to a homeomorphism of $\Psi_{c-a}^{-1}(D_a)$
      onto $D_a$, whose inverse maps $y$ to the point at level $c$ of the
      calibrated ray starting at $y$.
\item If $K=J^+(y)\cap T^{-1}([T(y),c])\subset O$, then $\cR_c(y)$ is a
      nonempty compact cellular subset of $\Lambda_c$; in particular it is
      connected and cell-like.
\item If $\Lambda_{c'}\subset O$ for some $c'\in\R$, then for every
      $s'\ge0$ the restriction $\Psi_{s'}:\Lambda_{c'}\to\Lambda_{c'-s'}$
      is a uniform limit of diffeomorphisms, hence a near-homeomorphism, and
      its fibres are cellular. This holds for every $c'$ if $O=M$.
\end{enumerate}
\end{theorem}

By Corollary~\ref{cor:cauchyline}, the assumption of (4) implies the
existence of a $b$-calibrated timelike line; under timelike completeness and
TCC the spacetime then splits. Statements (1) and (2) hold without any
assumption. In (2), neither the openness of $D_a$ in $\Lambda_a$ nor a
Lipschitz bound for the inverse is asserted.

The proof of (3) and (4) approximates $\cF$ by smooth vector fields. Norms
and convergence of vectors refer to the auxiliary metric, or to the
Euclidean norm in a chart.

\begin{lemma}[Smooth approximation of the normalized cones]\label{lem:smooth}
Let $A\subset O$ be compact. There are smooth future-directed timelike
vector fields $X_j$ on $M$, with $\dd T(X_j)=1$ and uniformly bounded on a
neighbourhood of $A$, such that
\begin{equation}\label{eq:graphlimit}
 x_j\to x\in A,\quad X_j(x_j)\to v\qquad\Longrightarrow\qquad v\in\cF_x .
\end{equation}
Consequently, for every $x\in A$ and $\epsilon>0$ there are $j_0$ and a
neighbourhood $W$ of $x$ such that, in a chart around $x$,
$X_j(x')\in\cF_x+\epsilon\overline B$ for all $j\ge j_0$ and $x'\in W$.
\end{lemma}

\begin{proof}
Fix compact sets $A'$ and $A''$ with $A\subset\operatorname{int}A'$,
$A'\subset\operatorname{int}A''$ and $A''\subset O$, finitely many charts
$(W_l,\varphi_l)$ with $W_l\subset O$, together with open sets
$W'_l\Subset W_l$ covering $A''$, and the smooth future timelike vector
field $Z=\nabla T/g(\nabla T,\nabla T)$, which satisfies $\dd T(Z)=1$. For
each $j$ choose a finite cover of $A'$ by open sets $W_{ij}\subset A''$ of
auxiliary diameter less than $1/j$, each contained in one of the sets
$W'_l$, and smooth functions $\phi_{ij}\ge0$ with compact support in
$W_{ij}$ and $\sum_i\phi_{ij}=1$ on $A'$. Choose $x_{ij}\in W_{ij}$ and
$v_{ij}\in\cF_{x_{ij}}$, and let $V_{ij}$ be the vector field on $W_{ij}$
whose components in the corresponding chart are constant and equal to
those of $v_{ij}$. The vectors of $\cF_x$, $x\in A''$, form a compact family
of future timelike vectors (Lemma~\ref{lem:F}), and the metric and the
derivatives of the transition maps are uniformly continuous on the
compact sets $\overline{W'_l}$; hence, for $j$ large, every $V_{ij}$ is
future timelike with $\dd T(V_{ij})\ge1/2$ on $W_{ij}$. Then
$Y_j=\sum_i\phi_{ij}V_{ij}$ is smooth on $M$, and future timelike with
$\dd T(Y_j)\ge1/2$ on $A'$. With a fixed smooth function $\chi$ equal to
one near $A$ and supported in $\operatorname{int}A'$ we put
\[
 X_j=\chi\,\frac{Y_j}{\dd T(Y_j)}+(1-\chi)Z .
\]
It is smooth and future timelike, as a convex combination of future
timelike vectors, it satisfies $\dd T(X_j)=1$, and it is uniformly bounded
near $A$.

Let $x_j\to x\in A$ with $X_j(x_j)\to v$. For $j$ large $\chi(x_j)=1$, and
in a chart $(W_l,\varphi_l)$ with $x\in W'_l$ every vector $V_{ij}(x_j)$ with $\phi_{ij}(x_j)\ne0$
differs from $v_{ij}\in\cF_{x_{ij}}$ by a quantity tending to zero
uniformly in $i$, since $d_a(x_{ij},x_j)<1/j$. As these $x_{ij}$ converge
to $x$, the upper semicontinuity of $\cF$ (Lemma~\ref{lem:F}) gives, for
every $\epsilon>0$ and $j$ large, $V_{ij}(x_j)\in\cF_x+\epsilon\overline B$,
hence $Y_j(x_j)\in\cF_x+\epsilon\overline B$ by convexity. Moreover
$\dd T(Y_j(x_j))\to1$, since $\dd T(v_{ij})=1$ at $x_{ij}$. Hence
$v\in\cF_x+\epsilon\overline B$ for every $\epsilon>0$, that is,
$v\in\cF_x$. The last statement follows from \eqref{eq:graphlimit} by
contradiction, using the uniform bound.
\end{proof}

If $X$ is a smooth future-directed timelike vector field with
$\dd T(X)=1$, its maximal integral curves are defined on $\R$: a maximal
integral curve defined only up to a finite time leaves every compact set,
so it would be an inextendible causal curve along which $T$ stays bounded,
while such a curve meets every level of $T$. Hence the flow $\phi_t$ of $X$
is complete, maps $\Lambda_c$ onto $\Lambda_{c+t}$, and
$\phi_{a-c}|_{\Lambda_c}:\Lambda_c\to\Lambda_a$ is a diffeomorphism, with
inverse $\phi_{c-a}|_{\Lambda_a}$.

\begin{proof}[Proof of Theorem~\ref{thm:fibres}]
(1) If $\Psi_s(z)=y$, then $T(z)=c$, and the trajectory of $\Psi$ from $z$,
reversed, is a future-directed $C^b$-causal curve from $y$ to $z$, so
$z\in\cR_c(y)$. Conversely, if $\eta$ is a future-directed $C^b$-causal
curve in $O$ from $y$ to $z\in\Lambda_c$, its reversal, parametrized by the
drop of $T$, is a past-directed $C^b$-causal curve from $z$, which by
Theorem~\ref{thm:unique}(1) coincides with $u\mapsto\Psi_u(z)$ on $[0,s]$;
hence $\Psi_s(z)=y$. By $T$-normalization, $\Psi_{s'}$ maps
$\Lambda_{c'}\cap O$ into $\Lambda_{c'-s'}\cap O$; if
$y'\in\Lambda_{c'-s'}\cap O$, a calibrated ray starting at $y'$ lies in $O$
and meets $\Lambda_{c'}$, and its point there is mapped to $y'$ by
$\Psi_{s'}$ (Theorem~\ref{thm:regular}(2)).

(2) Let $y\notin\cS$ and $z\in\cR_c(y)$, so that $\Psi_s(z)=y$. If
$\Psi_u(z)\in\cS$ for some $u\in[0,s]$, then
$y=\Psi_{s-u}(\Psi_u(z))\in\cS$, by Theorem~\ref{thm:unique}(2); hence
$s\le s_0(z)$, and by Theorem~\ref{thm:regular}(3) a calibrated ray starting
at $y$ passes through $z$. Only one calibrated ray starts at the regular
point $y$ \cite[Proposition~5.19]{minguzzi26e}, so $z$ is its point at level
$c$; conversely, this point belongs to $\cR_c(y)$ by (1) and
Theorem~\ref{thm:regular}(2). The points of a future-directed $C^b$-causal
curve $\eta$ starting at $y$ belong to the sets $\cR_{c''}(y)$,
$c''\ge T(y)$, hence to the calibrated ray. Let now $y\in\cS$ and $s>0$. As
in (1), the points at level $c$ of the calibrated rays starting at $y$
belong to $\cR_c(y)$, and at least two distinct calibrated rays start at
$y$ \cite[Proposition~5.19]{minguzzi26e}. If they met at a point $w$ at level
$c$, then by \cite[Proposition~5.6]{minguzzi26e}, applied at $w$, which is an
interior point of both, their parts after $w$ would coincide up to
reparametrization; being geodesics with a common tangent direction at $w$,
the two rays would coincide. Hence they meet $\Lambda_c$ at two distinct
points. The characterization of $\cS$ follows, since $\Psi_{s'}^{-1}(y)$
is nonempty by (1). Finally, by what precedes and (1), $\Psi_{c-a}$ maps
$\Psi_{c-a}^{-1}(D_a)$ continuously and bijectively onto $D_a$, with
inverse $y\mapsto\gamma_y(t_y)$, where $\gamma_y$ is the calibrated ray
starting at $y$, parametrized by proper time, and $T(\gamma_y(t_y))=c$. Let
$y_j\to y$ in $D_a$. By \cite[Theorem~5.12(v)]{minguzzi26e}, applied to the
compact set $\{y\}\cup\{y_j\}_j$, the initial unit tangents of the rays
$\gamma_{y_j}$ lie in a compact set, and their limit points are initial
tangents of calibrated rays starting at $y$, hence equal to that of
$\gamma_y$. By continuous dependence of geodesics on their initial data,
$\gamma_{y_j}\to\gamma_y$ uniformly on compact parameter intervals, and,
as $T\circ\gamma_y$ has positive derivative, $t_{y_j}\to t_y$ and
$\gamma_{y_j}(t_{y_j})\to\gamma_y(t_y)$. Thus the inverse is continuous.

(3) Let $a=T(y)$, $R=\cR_c(y)$, and assume $c>a$, the case $c=a$ being
trivial. By (1), $R$ is nonempty; it is closed in $\Lambda_c\cap O$, by the
continuity of $\Psi$, and it is contained in the compact set
$K\cap\Lambda_c\subset O$, so it is compact.

Choose a closed coordinate $n$-ball $B\subset\Lambda_a$ centred at $y$ such
that the compact set $K_B=J^+(B)\cap T^{-1}([a,c])$ is contained in $O$.
This is possible: otherwise there would be closed coordinate balls
$B_k\subset\Lambda_a$ shrinking to $y$ and points
$x_k\in J^+(B_k)\cap T^{-1}([a,c])\setminus O$, contained in the compact set
$J^+(B_1)\cap T^{-1}([a,c])$; a limit point $x$ would satisfy $x\in J^+(y)$,
by the closedness of the causal relation, $T(x)\in[a,c]$ and $x\notin O$,
as $O$ is open, contradicting $K\subset O$. The compact set
$K_-=J^-(R)\cap T^{-1}([a,c])$ is contained in $O$ by fact~1. Apply
Lemma~\ref{lem:smooth} to $A=K_B\cup K_-$, and let
$h_j=\phi^j_{a-c}|_{\Lambda_c}:\Lambda_c\to\Lambda_a$ be the past slice
diffeomorphisms of the fields $X_j$.

We first show that $\sup_{z\in R}d_a(h_j(z),y)\to0$. Let $z_j\in R$. The
integral curve of $X_j$ from $h_j(z_j)$ to $z_j$, parametrized by $T$, is
causal and ends in $R$, so it lies in $K_-$. These curves are uniformly
Lipschitz, and by the Arzel\`a--Ascoli theorem a subsequence converges
uniformly to a curve from a point of $\Lambda_a$ to a point $z\in R$. The
argument of the paragraph {\em The limiting inclusion} in the proof of
Proposition~\ref{prop:chains}, with the last statement of
Lemma~\ref{lem:smooth} in place of the upper semicontinuity of $\cF$,
shows that the limit has velocity in $\cF$ almost everywhere. Reversed, it
is a past-directed $C^b$-causal curve from $z$ parametrized by the drop of
$T$, hence, by Theorem~\ref{thm:unique}(1), the trajectory of $\Psi$ from
$z$, whose point on $\Lambda_a$ is $\Psi_{c-a}(z)=y$. Thus
$h_j(z_j)\to y$ along the subsequence, and the claim follows.

Let $B_j\subset B$ be closed coordinate balls centred at $y$ whose radii
tend to zero and exceed the maximal coordinate distance of $h_j(R)$ from
$y$, and put $C_j=h_j^{-1}(B_j)$, a closed $n$-cell in $\Lambda_c$ with
$R\subset\operatorname{int}C_j$. For every neighbourhood $U$ of $R$ in
$\Lambda_c$, $C_j\subset U$ for $j$ large. Otherwise, along a subsequence,
there are $z_j\in C_j\setminus U$; the integral curve of $X_j$ from
$h_j(z_j)\in B_j$ to $z_j$ is causal and starts in $B$, so it lies in
$K_B$, and as above a further subsequence converges to a future-directed
$C^b$-causal curve from $y$ to a point $z\in\Lambda_c\setminus U$. Then
$z\in\cR_c(y)=R\subset U$, a contradiction. Choosing inductively
$C_{j_{k+1}}\subset\operatorname{int}C_{j_k}$ within the
$1/k$-neighbourhood of $R$, we obtain $R=\bigcap_kC_{j_k}$, so that $R$ is
cellular.

(4) If $\Lambda_{c'}\subset O$, then $T^{-1}((-\infty,c'])\subset
J^-(\Lambda_{c'})\subset O$ by fact~1, since a future inextendible causal
curve from a point $x$ with $T(x)\le c'$ meets $\Lambda_{c'}$. Apply
Lemma~\ref{lem:smooth} to the compact set $A=T^{-1}([c'-s',c'])\subset O$, and let $h_j:\Lambda_{c'}\to\Lambda_{c'-s'}$ be the
past slice diffeomorphisms of the fields $X_j$. If $z_j\to z$ in
$\Lambda_{c'}$, the argument of (3) shows that every subsequence of
$h_j(z_j)$ has a further subsequence converging to
$\Psi_{s'}(z)$. Thus $h_j$ converges continuously, hence uniformly on the
compact set $\Lambda_{c'}$, to the continuous map $\Psi_{s'}$. Its fibres
are cellular by (3), whose assumption holds, as
$J^+(y)\cap T^{-1}([T(y),c'])\subset A$ for $y\in\Lambda_{c'-s'}$.
\end{proof}

\begin{remark}[Connectedness through Kneser's theorem]\label{rem:kneser}
The connectedness of the fibres, under the assumption of
Theorem~\ref{thm:fibres}(3), can also be derived without smooth
approximation, from a Kneser-type theorem on the solution sets of
differential inclusions with upper semicontinuous compact convex
right-hand sides \cite[Theorem~1, Remarks~1 and~3, Lemma~3]{KK}, applied
on short temporal steps and propagated by the semigroup law.
\end{remark}

\begin{remark}[An example of fibres]\label{rem:fibres}
In the model of Remark~\ref{rem:merge}, with $T=t$, the fibre of $\Psi_s$
over a point $(t_0,0)$ of the singular axis is the segment
$\{(t_0+s,z):|z|\le(a/A)s\}$, a cellular set, while the fibre over a
regular point $(t_0,z_0)$, $z_0\ne0$, is the single point
$(t_0+s,z_0+(a/A)s\operatorname{sgn}z_0)$ of the calibrated ray starting at
$(t_0,z_0)$, in accordance with Theorem~\ref{thm:fibres}.
\end{remark}

\begin{remark}[Finite speed of dilation]\label{rem:speed}
Let $y\in O$ and $s\ge0$, put $\Lambda=T^{-1}(T(y)+s)$, and assume, as in
Theorem~\ref{thm:fibres}(3), that $K=J^+(y)\cap T^{-1}([T(y),T(y)+s])$ is
contained in $O$. By Lemma~\ref{lem:F} there are $m_0,L>0$ such that
$g(v,v)\le-m_0$ and $\norma v\le L$ for every $v\in\cF_x$, $x\in K$. Fix
$\epsilon\in(0,m_0)$ so small that $g_\epsilon=g+\epsilon\,\dd T\otimes\dd T$
is a Lorentzian metric on a neighbourhood $N$ of $K$, time oriented by
$\dd T$. Since $g_\epsilon(w,w)\ge g(w,w)$, with
$g_\epsilon(w,w)=\epsilon\,\dd T(w)^2>0$ for nonzero $g$-null vectors $w$, the
causal cones of $g_\epsilon$ lie, apart from the origin, in the open timelike
cones of $g$; and since $g_\epsilon(v,v)\le-m_0+\epsilon<0$ for
$v\in\cF_x$, the cone $C^b_x$ lies in the timelike cone of $g_\epsilon$ for
$x\in K$. Every future-directed $C^b$-causal curve from $y$ to $\Lambda$
lies in $K$; parametrized by $T$, it is $g_\epsilon$-timelike almost
everywhere and has auxiliary speed at most $L$. Hence, by
Theorem~\ref{thm:fibres}(1),
\[
 \Psi_s^{-1}(y)\subset J^+_{g_\epsilon}(y)\cap\Lambda,\qquad
 \operatorname{diam}_a\Psi_s^{-1}(y)\le2Ls,
\]
where $J^+_{g_\epsilon}(y)$ denotes the causal future of $y$ for $g_\epsilon$
within $N$, which for $s>0$ meets $\Lambda$ inside $I^+(y)$.

In plain terms, the region of the later slice which the past semi-flow
collapses onto a singular point spreads out from that point at a finite
speed, strictly lower than the speed of light $c$. Its boundary moves like
a front of massive observers, never like a light front, and its size grows
at most linearly in time. In the model of Remark~\ref{rem:fibres} the fibre
over a singular point is a segment of half-width $(a/A)s$, so its endpoints
move with speed $a/A=a/\sqrt{1+a^2}<1$ in units with $c=1$. Speeds are
measured here with the time function $T$ and the auxiliary metric; the
inclusion in the causal future of $y$ for the metric $g_\epsilon$, whose
cones are strictly narrower than the light cones, is the invariant
formulation of this statement.
\end{remark}

\subsection{Forward calibrated maps}

For $\delta>0$ put
\[
 A_\delta=\{p\in O:s_0(p)\ge\delta\}.
\]
By Proposition~\ref{prop:entry}, $A_\delta$ is closed in $O$, and it
consists of regular points. By Theorem~\ref{thm:regular}(3), every
$p\in A_\delta$ lies on a calibrated ray which starts at a point $q$ with
$T(p)-T(q)\ge\delta$; in this sense the calibrated past extension of $p$ has
temporal length at least $\delta$.

\begin{theorem}[Lipschitz differential]\label{thm:gradient}
For every $\delta>0$ the map $p\mapsto\dd b_p$ is locally Lipschitz on
$A_\delta$, in local trivializations of $T^*M$ and for the relative
topology of $A_\delta$.
\end{theorem}

\begin{proof}
Fix a coordinate ball $W$ with compact closure in $O$ and a compact
neighbourhood $W'\subset O$ of $\overline W$. The unit tangents of
calibrated rays at points of $W'$ form a compact family of future timelike
vectors (fact~2). Choose $r>0$ so small that, for every $p\in\overline W$
and every such tangent $u$ at $p$, the geodesic segment
$s\mapsto\exp_p(-su)$, $0\le s\le r$, lies in $W'$, its temporal drop is
less than $\delta$, and the coordinate ball of radius $\rho$ around $p$ lies
in the chronological future of $\exp_p(-ru)$, where the Lorentzian distance
from this point is smooth, with coordinate Hessian bounded by a constant $C$
independent of $p$ and $u$. This is
possible because the pairs $(p,u)$ range over a compact set and, for fixed
$r$, the relevant pairs of points stay in a compact subset of the set of
chronologically related pairs within convex normal neighbourhoods, where
the distance is smooth.

Let $p\in A_\delta\cap W$ and $u_p=-\nabla b(p)$. By
Theorem~\ref{thm:regular}(3) with $s=\delta$, a calibrated ray passes
through $p$, with tangent $u_p$ there, and its segment of temporal drop
$\delta$ before $p$ is calibrated. As the drop of
$s\mapsto\exp_p(-su_p)$, $0\le s\le r$, is less than $\delta$, the point
$q_p=\exp_p(-ru_p)$ lies on that segment, and
$\ell_p=b(q_p)+d(q_p,\cdot)$ is a smooth lower support of $b$ at $p$, by
the rushing inequality and calibration. As $b$ is differentiable at $p$,
$\dd(\ell_p)_p=\dd b_p$, and therefore
\begin{equation}\label{eq:loweruniform}
 b(y)\ge b(p)+\dd b_p(y-p)-C|y-p|^2
 \qquad(p\in A_\delta\cap W,\ |y-p|\le\rho).
\end{equation}
By \eqref{eq:semiconc}, after shrinking $\rho$, with a uniform $A\ge0$,
\begin{equation}\label{eq:upperuniform}
 b(y)\le b(p)+\dd b_p(y-p)+A|y-p|^2 .
\end{equation}
The points $y$ in these inequalities need not belong to $A_\delta$.

Let $p,q\in A_\delta\cap W$ with $0<\varrho=|q-p|\le\rho/2$, and let $e$ be
a coordinate unit vector. Evaluating \eqref{eq:upperuniform} at $p$ and
\eqref{eq:loweruniform} at $q$ on the point $y=q+\varrho e$ gives
\[
 \varrho(\dd b_q-\dd b_p)e
 \le b(p)+\dd b_p(q-p)-b(q)+A|q-p+\varrho e|^2+C\varrho^2 .
\]
By \eqref{eq:loweruniform} at $p$, evaluated at $q$, the first three terms
on the right are at most $C\varrho^2$. Hence
$(\dd b_q-\dd b_p)e\le(4A+2C)\varrho$, and taking the supremum over $e$
gives $|\dd b_q-\dd b_p|\le(4A+2C)|q-p|$. Larger separations in $W$ are
handled by the boundedness of $\dd b$.
\end{proof}

The next result makes Theorems~\ref{thm:support} and~\ref{thm:gradient}
quantitative: up to constants, the temporal length of the calibrated past
extension of a point is the reciprocal of the constant of the best
quadratic lower support of $b$ on the time slice through that point. This
parallels the fact that the Hessian of the Lorentzian distance from a point
$q$, restricted to a spacelike slice, is of the order of the reciprocal of
the distance from $q$.

\begin{theorem}[Entry time and semiconvexity on time slices]\label{thm:quant}
Let $W\subset O$ be the domain of a chart $(\tau,\xi)\in\R\times\R^n$ with
$\tau=T|_W$, and write $b$ in these coordinates. For $p\in W$ with
coordinates $(\tau_p,\xi_p)$ and $\rho>0$, let $C_\rho(p)\in[0,\infty]$ be
the infimum of the constants $C\ge0$ for which there is a covector $P$ with
\begin{equation}\label{eq:sliceprox}
 b(\tau_p,\xi)\ge b(\tau_p,\xi_p)+P(\xi-\xi_p)-C|\xi-\xi_p|^2
 \qquad(|\xi-\xi_p|\le\rho).
\end{equation}
Let $K\subset W$ be compact. There are $\rho,\delta_0,c_1,c_2,c_3>0$ such
that, for every $p\in K$,
\begin{equation}\label{eq:quant}
 c_1\min\{1,C_\rho(p)^{-1}\}\le s_0(p),\qquad
 C_\rho(p)\le\frac{c_2}{\min\{s_0(p),\delta_0\}},
\end{equation}
with $1/0=\infty$ and $1/\infty=0$, and
\begin{equation}\label{eq:quantlip}
 |\dd b_q-\dd b_p|\le\frac{c_3}{\delta}\,|q-p|
 \qquad(p,q\in A_\delta\cap K,\ 0<\delta\le\delta_0).
\end{equation}
In particular, $\cE\cap K=\{p\in K:C_\rho(p)=\infty\}$, and
$c_1/C_\rho(p)\le s_0(p)\le c_2/C_\rho(p)$ for every $p\in K$ with
$s_0(p)\le\delta_0$ and $C_\rho(p)\ge1$.
\end{theorem}

\begin{proof}
Choose $\rho_1>0$ and a compact set $K'\subset W$ containing the closed
coordinate $\rho_1$-neighbourhood of $K$. Let $\Lambda$ bound the
coordinate norms of $\dd b$ at the differentiability points of $K'$, let
\eqref{eq:semiconc} hold with constant $A$ for $x,y\in K'$ with
$|y-x|\le\rho_1$, let $\mathcal U$ be the compact set of the unit tangents
of calibrated rays at points of $K'$ (fact~2), and put
$m=\frac12\min_{\mathcal U}\dd T$ and $M_T=\max_{\mathcal U}\dd T$.

{\em The first inequality.} Apply Lemma~\ref{lem:cap} to the levels of
$T$, parametrized by $z_p(\xi)=(\tau_p,\xi_p+\xi)$, $p\in K$, and to
$u\in U_p\subset\mathcal U$. By the uniformity statement of
Lemma~\ref{lem:cap}, its constants $r_0,c,C_*$ can be chosen independent
of $p$ and $u$, and, decreasing $r_0$, we may assume
$T(p)-T(\exp_p(-ru))\ge mr$ for $0<r\le r_0$. Let $\rho\le\rho_1$, and let
$p\in K$ with $C_\rho(p)<\infty$. For $C'>C_\rho(p)$, inequality
\eqref{eq:sliceprox} holds with $C'$ in place of $C$; put
$r=\min\{r_0,c/C',\rho/C_*\}$. Then $D_r\subset\{|\xi|\le C_*r\}\subset
\{|\xi|\le\rho\}$ and $c/r\ge C'$, and the proof of (4)$\Rightarrow$(1) in
Theorem~\ref{thm:support} shows that $p$ is an interior point of a
calibrated ray starting at $\exp_p(-ru)$, where $U_p=\{u\}$. By
\eqref{eq:s0}, $s_0(p)\ge mr$. Letting $C'\downarrow C_\rho(p)$ gives
$s_0(p)\ge m\min\{r_0,\rho/C_*,c/C_\rho(p)\}$, which is the first
inequality in \eqref{eq:quant}, with $c_1=m\min\{r_0,\rho/C_*,c\}$. If
$C_\rho(p)=\infty$ there is nothing to prove.

{\em The second inequality and the Lipschitz bound.} Put
$\kappa_1=1/(2M_T)$. For $\delta_0$ small and $0<r\le\kappa_1\delta_0$, every
geodesic segment $s\mapsto\exp_x(-su)$, $0\le s\le r$, with $x\in K$ and
$u\in\mathcal U$, has temporal drop less than $r/\kappa_1$. Let
$0<\delta\le\delta_0$, let $p\in A_\delta\cap K$, and put $r=\kappa_1\delta$
and $q_p=\exp_p(-ru_p)$. As in the proof of Theorem~\ref{thm:gradient},
$q_p$ lies on the calibrated segment of temporal drop $\delta$ before $p$,
so that $\ell_p=b(q_p)+d(q_p,\cdot)\le b$ on the chronological future of
$q_p$ near $p$, with equality at $p$, and $\dd(\ell_p)_p=\dd b_p$. In normal
coordinates centred at $q_p$, $d(q_p,\cdot)=W^{1/2}$ on the chronological
future of $q_p$, with $W(y)=-g_{q_p}(\exp_{q_p}^{-1}y,\exp_{q_p}^{-1}y)$.
Since $\exp_{q_p}^{-1}p=r\,\dot\gamma$ with $\dot\gamma$ in a compact
family of future unit timelike vectors, there is $\kappa_2>0$, independent
of $p$ and $\delta$, such that on the coordinate ball $B_p$ of radius
$\kappa_2r$ around $p$ we have $W\ge r^2/4$, $|DW|=O(r)$ and $|D^2W|=O(1)$,
and $\exp_{q_p}^{-1}y$ is future timelike. Hence $B_p$ lies in the
chronological future of $q_p$, and
\[
 D^2d(q_p,\cdot)=\frac{D^2W}{2W^{1/2}}-\frac{DW\otimes DW}{4W^{3/2}}
 =O(1/r)
\]
on $B_p$, uniformly in $p$ and $\delta$. Taylor's formula gives a constant
$\kappa_3$, independent of $p$ and $\delta$, with
\begin{equation}\label{eq:lowerdelta}
 b(y)\ge b(p)+\dd b_p(y-p)-\frac{\kappa_3}{r}|y-p|^2\qquad(y\in B_p).
\end{equation}
For $y=(\tau_p,\xi)$ with $\kappa_2r\le|\xi-\xi_p|\le\rho$, the Lipschitz
bound gives $|b(y)-b(p)-\dd b_p(y-p)|\le2\Lambda|\xi-\xi_p|\le
(2\Lambda/(\kappa_2r))|\xi-\xi_p|^2$. Therefore \eqref{eq:sliceprox} holds
with $C=\max\{\kappa_3,2\Lambda/\kappa_2\}/r$, and $C_\rho(p)\le c_2/\delta$
with $c_2=\max\{\kappa_3,2\Lambda/\kappa_2\}/\kappa_1$. If $s_0(p)>0$, this
applies with $\delta=\min\{s_0(p),\delta_0\}$ and gives the second
inequality in \eqref{eq:quant}, which is trivial if $s_0(p)=0$.

Let now $p,q\in A_\delta\cap K$ and $\varrho=|q-p|$. If
$\varrho\le\kappa_2r/2$, the argument of the proof of
Theorem~\ref{thm:gradient}, with \eqref{eq:lowerdelta} in place of
\eqref{eq:loweruniform}, gives
$|\dd b_q-\dd b_p|\le(4A+2\kappa_3/r)\varrho$. Otherwise
$|\dd b_q-\dd b_p|\le2\Lambda\le(4\Lambda/(\kappa_2r))\varrho$. As
$r=\kappa_1\delta$ and $\delta\le\delta_0$, this proves \eqref{eq:quantlip}
with $c_3=\max\{4A\delta_0+2\kappa_3/\kappa_1,4\Lambda/(\kappa_1\kappa_2)\}$,
where $\delta_0$ is taken so small that $\kappa_1\kappa_2\delta_0\le\rho$.

Finally, by \eqref{eq:quant}, $s_0(p)=0$ implies $C_\rho(p)=\infty$, and
$C_\rho(p)=\infty$ implies $\min\{s_0(p),\delta_0\}=0$, that is,
$s_0(p)=0$. The two-sided estimate follows directly from \eqref{eq:quant}.
\end{proof}

For a regular point $p$ and $s\ge0$, let $P_s(p)$ be the point at level
$T(p)+s$ of the calibrated ray starting at $p$.

\begin{corollary}[Forward calibrated maps and the line flow]\label{cor:forward}
For every $\delta>0$ the map $(s,p)\mapsto P_s(p)$ is jointly locally
Lipschitz on $[0,\infty)\times A_\delta$. For every $s\ge0$, $P_s$ maps
$A_\delta$ bijectively onto $A_{\delta+s}$, with inverse $\Psi_s$, so that
$P_s:A_\delta\to A_{\delta+s}$ is locally bi-Lipschitz. On $\mathcal L_b$ the
maps $G_t=P_t$, $t\ge0$, and $G_t=\Psi_{-t}$, $t\le0$, form a jointly
locally Lipschitz one-parameter group with $T(G_t(p))=T(p)+t$, whose orbits
are the $b$-calibrated timelike lines.
\end{corollary}

\begin{proof}
By Theorem~\ref{thm:gradient}, $p\mapsto u_p=-\nabla b(p)$ is locally
Lipschitz on $A_\delta$. The geodesic flow is smooth, and the proper time at
which the geodesic with initial data $(p,u_p)$ meets the level $T(p)+s$
depends smoothly on $(p,u_p,s)$ near each crossing, by the implicit function
theorem, since $\dd T$ is positive on future timelike vectors. Hence $P_s(p)$
is jointly locally Lipschitz on $[0,\infty)\times A_\delta$.

By Corollary~\ref{cor:structure}(1), $s_0(P_s(p))=s_0(p)+s$, so that
$P_s(A_\delta)\subset A_{\delta+s}$, and $\Psi_s(P_s(p))=p$ by
Theorem~\ref{thm:regular}(2). Conversely, if $z\in A_{\delta+s}$, then
$p=\Psi_s(z)\in A_\delta$ by Proposition~\ref{prop:entry}, and by
Theorem~\ref{thm:regular}(3) the reversed trajectory from $p$ to $z$ is a
segment of a calibrated ray, which is the unique calibrated ray starting at
the regular point $p$; hence $P_s(p)=z$. The inverse $\Psi_s$ is locally
Lipschitz by Theorem~\ref{thm:unique}(2). Finally,
$\mathcal L_b\subset A_\delta$ for every $\delta$, and $P_s$ and $\Psi_s$
preserve $\mathcal L_b$, by Corollary~\ref{cor:structure}(1) and
Proposition~\ref{prop:entry}; the identities $\Psi_s\circ P_s=\mathrm{id}$
and $P_s\circ\Psi_s=\mathrm{id}$ on $\mathcal L_b$ give the group law, and
the orbits are the lines of Corollary~\ref{cor:structure}(2). Joint local
Lipschitz continuity holds for $t\ge0$ and for $t\le0$, and hence, by the
triangle inequality across $t=0$, on $\R$.
\end{proof}

Restricted to levels of $T$, Corollary~\ref{cor:forward} gives locally
bi-Lipschitz bijections $P_s:A_\delta\cap\Lambda_a\to A_{\delta+s}\cap
\Lambda_{a+s}$, with inverse $\Psi_s$. This sharpens the homeomorphism of
Theorem~\ref{thm:fibres}(2) on the sets where the entry time is bounded
away from zero. No Lipschitz bound for the inverse is asserted near regular
points of $\cE$, and no manifold structure is asserted for $A_\delta$ or
$\mathcal L_b$.

\section{Time duality: the cosmological time}\label{sec:cosmo}

In this section we prove Theorem~\ref{thm:cosmo}. Let $(M,g)$ have regular
cosmological time $\tau$. By \cite{andersson98}, $(M,g)$ is globally
hyperbolic (Proposition~2.2), $\tau$ is locally Lipschitz and locally
semiconvex (Theorem~1.2 and Proposition~3.1), and
\begin{equation}\label{eq:revlip}
 \tau(x)+d(x,y)\le\tau(y)\qquad(x\le y)
\end{equation}
(Proposition~2.1). Every $x\in M$ is the endpoint of a rest curve
(Proposition~2.3), and the final tangents of the rest curves ending at the
points of a compact set form a bounded set (Proposition~2.4).

\begin{lemma}\label{lem:dual}
Let $\hat M$ be $M$ with the opposite time orientation, so that its
Lorentzian distance is $\hat d(x,y)=d(y,x)$, and let $\hat b=-\tau$. Then
facts~1--5 of Section~\ref{sec:setting} hold for $\hat b$ on $\hat O=M$,
with $\hat U_x=-R_x$, except that \eqref{eq:support} holds only for
$0<\sigma<\tau(x)$.
\end{lemma}

\begin{proof}
Fact~1 is trivial. For fact~2, $\hat b$ is locally Lipschitz and locally
semiconcave, $R_x\ne\varnothing$, and the union of the
$R_x$, $x\in K$, over a compact set $K$ is bounded and closed in $TM$, as a
limit of rest curves is a rest curve, by the limit curve theorem, the upper
semicontinuity of the length and the continuity of $\tau$; hence it is
compact. For fact~3, let $u\in R_x$ be the final tangent of the rest curve
$\gamma$, let $0<\sigma<\tau(x)$ and $z=\gamma(\tau(x)-\sigma)
=\exp_x(-\sigma u)$. As $\tau(z)=\tau(x)-\sigma$, \eqref{eq:revlip} gives
$\tau(y)\ge\tau(x)-\sigma+d(z,y)$ for $y\gg z$, which is \eqref{eq:support}
for $\hat b$ at the vector $-u$, future directed in $\hat M$. Since $\gamma$
is maximizing and extends beyond $z$ to the past, $x$ is not a future cut
point of $z$, so $d(z,\cdot)$ is smooth near $x$, with differential
$-g_x(u,\cdot)$ at $x$, as in \cite[Theorem~5.12(iii)]{minguzzi26e}. For
fact~4, if $\tau$ is differentiable at $x$, these smooth lower supports
touch $\tau$ at $x$, so $\dd\tau_x=-g_x(u,\cdot)$ for every $u\in R_x$.
Hence $R_x=\{-\nabla\tau(x)\}$, $g(\nabla\tau,\nabla\tau)=-1$, and
$-\nabla\hat b(x)=\nabla\tau(x)\in\hat U_x$. Fact~5 is \eqref{eq:revlip}.
\end{proof}

\begin{proof}[Proof of Theorem~\ref{thm:cosmo}]
The proofs of Sections~\ref{sec:strength}--\ref{sec:unique}, except
Section~\ref{sec:consequences}, use the
standing assumptions only through facts~1--5 and global hyperbolicity, and
they use \eqref{eq:support} only with the fixed small $\sigma$ of
Lemma~\ref{lem:gauss}, chosen uniformly on compact sets. Since $\tau$ has a
positive lower bound on compact sets, Lemma~\ref{lem:dual} allows us to
apply them to $\hat b$ in $\hat M$. Reversing the time orientation again,
we get (1), (2) and the first part of (3).

Corollary~\ref{cor:geometry}(1)--(2) follow as in Section~\ref{sec:consequences},
except for the measure statement. For the latter, by
Lemmas~\ref{lem:strength} and~\ref{lem:dual} the dimension of the affine
span of $R_x$ equals that of the subdifferential of $\tau$ at $x$, so the
set of the points at which it is at least $k$ is countably
$\mathcal H^{n+1-k}$-rectifiable \cite{alberti92}, $\tau$ being locally
semiconvex. With this in place of the theory of horizons, the proof of
\cite[Theorem~5.21(c)]{minguzzi26e} applies verbatim: it uses only the
quadratic supports of fact~3, which hold uniformly on compact sets, and
the compactness and closedness of the sets $R_x$. This proves the rest of
(3).

For (4), by (3) a singular point in $J^-(\Sigma)$ would give a singular
point on $\Sigma$, so $\tau$ is differentiable on $J^-(\Sigma)$, and a locally semiconvex function which is differentiable
on an open set is $C^1$ there. Let $K\subset I^-(\Sigma)$ be compact. The
integral curves of the continuous unit vector field $-\nabla\tau$ starting
from the points of $K$ exist for a uniform proper time $\delta>0$ inside
$I^-(\Sigma)$, and $\tau$ increases at unit rate along them, since
$\dd\tau(-\nabla\tau)=1$. Thus the concatenation of a rest curve ending at
$x\in K$ with the integral curve from $x$ to its point $y$ at proper time
$\delta$ has length $\tau(y)$: it is a rest curve of $y$, hence a geodesic,
and $x$ is an interior point of it. By \eqref{eq:revlip},
$\tau\le\tau(y)-d(\cdot,y)$ near $x$, with equality at $x$, and, as above,
$d(\cdot,y)$ is smooth near $x$, with second derivatives bounded uniformly
in $x\in K$, the points $y$ ranging in a compact subset of $I^+(K)$ at
proper time $\delta$ along geodesics with tangents in a compact set. Hence
$\tau$ is also locally semiconcave on $I^-(\Sigma)$, and a function which
is both locally semiconvex and locally semiconcave is
$C^{1,1}_{\mathrm{loc}}$.
\end{proof}

\begin{remark}[Two arriving rest curves]\label{rem:midpoint}
If $R_x=\{u_+,u_-\}$, then $V_\tau(x)=(u_++u_-)/2$, since
$\normL{\lambda u_++(1-\lambda)u_-}^2=1+2\lambda(1-\lambda)
(\cosh\angle(u_+,u_-)-1)$ is maximal at $\lambda=1/2$. Thus, at almost
every time at which it lies in the stratum where exactly two rest curves
arrive, the singular curve of Theorem~\ref{thm:cosmo}(3) moves in the
direction of the mean $u_++u_-$ of the two arriving velocities. The same
holds for $b$ with the two departing velocities of the calibrated rays.
\end{remark}

The results of Sections~\ref{sec:rays} and~\ref{sec:fibres} do not dualize as
directly, because the rest curves have finite length: in particular the
calibrated lines have no counterpart, and every forward trajectory of
$\Psi^\tau$ through a regular point runs along the rest curves only as
long as it stays in the regular region.

\section{Conclusions}\label{sec:conclusions}

We have shown that the singularities of the future Cauchy Busemann function
have a rigid causal structure. They propagate toward the past along the
causal curves of the calibrated cone structure, which are timelike, and
their strength decays at most exponentially along these curves. The curves
are unique: they are the trajectories of a locally Lipschitz past semi-flow,
which follows the calibrated rays backward in the regular region, is
eventually captured by the singular set away from calibrated lines, and
merges trajectories there. Toward the future the same curves branch, and
they branch exactly at the singular points. The singular set thus behaves
as a past-rooted web of timelike creases in the field of maximally aging
observers.

The future cosmological development decomposes accordingly into the
calibrated timelike lines, the set $\cE$ of initial points, which lies
between $\cS$ and its closure, coincides with the analytic cut locus of $b$
and has measure zero on spacelike hypersurfaces, and the interiors of the
maximal calibrated rays, each of which starts on $\cE$. Away from $\cE$,
quantitatively on the sets $A_\delta$, the differential of $b$ is locally
Lipschitz, with constant of order $1/\delta$, and the forward calibrated
maps are locally bi-Lipschitz. The temporal length of the calibrated past
extension of a point is comparable to the reciprocal of the semiconvexity
defect of $b$ on the time slice through it. The
fibres of the semi-flow are single points over regular points, while over
singular points they are cellular sets with more than one point, whenever
the relevant compact future region lies in the development; as time
increases, these fibres spread out from the singular points at a speed
lower than the speed of light.

For Bartnik's conjecture, these results describe the shape that a
counterexample must have: a nonempty singular set rooted in the infinite
past, a past semi-flow which drives every point of the development into the
singular set, maximal calibrated rays all born on the closure of the
singular set, and a development which contains no Cauchy hypersurface,
whose past is infinitely deep and along which $b+d(\cdot,\Lambda)$ diverges at
past infinity. None of these statements uses a curvature condition.

The same analysis applies, with the time orientation reversed, to the
cosmological time of a spacetime with an initial singularity
(Theorem~\ref{thm:cosmo}). Its singular points, the events reached by more
than one maximal geodesic from the initial singularity, persist once
formed, and move toward the future along unique timelike curves, while the
singularities of the canonical Cauchy Busemann function persist toward the
past. The two singular sets are thus complementary: the first records the
crossings of the maximally aging observers coming from the initial
singularity, the second the branching of those going to the far future. In a cosmology with regular cosmological time the two
structures coexist with calibrated timelike lines of the canonical function
$b^{\textnormal{\textsc{m}}}$: eternal maximally aging observers, emanating
from the initial singularity, which are never captured by the singular set. A proof of the conjecture
along these lines must therefore bring the timelike convergence condition
into play on this structure, for instance through the focusing of the
maximal calibrated rays emanating from $\cE$, combined with the mean
convexity of the level sets of $b$ in the support sense established in
\cite{minguzzi26e}.

Several questions remain open. The singular set $\cS$ is countably
$n$-rectifiable \cite[Theorem~5.21(c)]{minguzzi26e}, but we do not know
whether $\cE$ is, or whether it has Hausdorff dimension at most $n$;
nullity on spacelike hypersurfaces does not imply this. The examples showing the
failure of lower semicontinuity of the entry time, the regularity of some
entry points, and the existence of spacelike intrinsic arcs are local
models, and it is not known whether these phenomena occur for Cauchy
Busemann functions of spacetimes with compact Cauchy hypersurfaces. Finally,
the regularity of the semi-flow and of the forward calibrated maps near
$\cE$, where the past extension of the calibrated rays degenerates, remains
to be understood.

\section*{Acknowledgments}

For the writing of this paper we had the assistance of AI tools from
Anthropic (Claude Opus 5.5) and OpenAI (GPT-6 Astra). This work was supported
in part by a grant of access to OpenAI models through the ChatGPT for Academic
Researchers program.

\appendix
\section{Flat affine models and the closedness issue}\label{sec:flat}

The superdifferential of a minimum of a compact family of affine functions
is given by the following elementary formula, of Danskin type, which we
prove directly.

\begin{lemma}[Compact families of affine functions]\label{lem:danskin}
Let $\mathcal A\subset\R\times(\R^m)^*$ be a nonempty compact set and
\[
 b(x)=\min_{(a,p)\in\mathcal A}\{a+p(x)\}\qquad(x\in\R^m).
\]
Let $\mathcal A(x)$ be the set of active pairs, those attaining the
minimum at $x$, and $P(x)=\{p:(a,p)\in\mathcal A(x)\}$. Then $P(x)$ is
nonempty and compact, and for every $v\in\R^m$
\begin{equation}\label{eq:danskin}
 b'(x;v)=\min_{p\in P(x)}p(v),\qquad D^+b(x)=\co P(x).
\end{equation}
\end{lemma}

\begin{proof}
The function $b$ is concave and Lipschitz, with constant
$\max_{\mathcal A}|p|$. As $b$ is continuous, $\mathcal A(x)$ is a nonempty
closed subset of $\mathcal A$, so $P(x)$ is compact and so is $\co P(x)$,
by Carath\'eodory's theorem. For $(a,p)\in\mathcal A(x)$ and $t>0$,
$b(x+tv)\le a+p(x+tv)=b(x)+t\,p(v)$. Conversely, choose
$(a_t,p_t)\in\mathcal A(x+tv)$. Then
$b(x+tv)-b(x)\ge a_t+p_t(x+tv)-a_t-p_t(x)=t\,p_t(v)$, so that
\[
 p_t(v)\le\frac{b(x+tv)-b(x)}t\le\min_{p\in P(x)}p(v).
\]
Every limit point $(a_*,p_*)$ of $(a_t,p_t)$ as $t\downarrow0$ belongs to
$\mathcal A$, and $a_*+p_*(x)=\lim_t[b(x+tv)-t\,p_t(v)]=b(x)$, so it is
active at $x$. Hence $\liminf_{t\downarrow0}p_t(v)\ge\min_{P(x)}p(v)$,
which proves the first identity in \eqref{eq:danskin}. Each $p\in P(x)$
belongs to $D^+b(x)$, as $a+p(\cdot)$ is an affine upper support touching
at $x$, and $D^+b(x)$ is convex; thus $\co P(x)\subset D^+b(x)$.
Conversely, if $Q\in D^+b(x)$, then $b'(x;v)\le Q(v)$ for every $v$, that
is, $Q(v)\ge\min_{p\in\co P(x)}p(v)$. If $Q$ were not in the compact convex
set $\co P(x)$, strict separation would give $v$ with
$Q(v)<\min_{p\in\co P(x)}p(v)$, a contradiction.
\end{proof}

\begin{corollary}[Monotonicity in affine models]\label{cor:flat}
Suppose, in Minkowski space, that $b$ is an infimum of affine functions
with past unit timelike gradients, with nonempty compact active slope sets
and locally compact slope bounds, and that \eqref{eq:superdiff} holds with
these active slopes. Then $\vartheta(x(t))$ is nondecreasing along every
past-directed $C^b$-causal curve, and $\vartheta(z)\ge\vartheta(x)$ along
every sphere step $x\mapsto z$. By Lemma~\ref{lem:danskin}, the hypotheses
hold for a minimum of affine functions with past unit timelike gradients
whose coefficient set is compact, for instance a minimum indexed by a
compact space with continuous coefficients; no finiteness or
distinct-gradient assumption is needed.
\end{corollary}

\begin{proof}
Every active affine function is an exact global upper support. Hence, for
a short timelike chord, the proof of Lemma~\ref{lem:initial} gives
$b(q)-b(y)\ge r\vartheta(q)$ with no remainder, and Lemma~\ref{lem:rigidity}
holds verbatim. In proper time $r\ge h$, so \eqref{eq:drop} yields
$\frac1h\int_t^{t+h}f\ge f(t)$, and Lemma~\ref{lem:average} applies with
$\omega=0$. Likewise, the two averaged support inequalities in the proof
of Proposition~\ref{prop:persistence} hold without remainder and give
$\Theta_x\le\Theta_z$.
\end{proof}

\begin{remark}[A calibrated ray can meet the closure of the singular set]\label{rem:closure}
The proof of Theorem~\ref{thm:invariance} does not require excluding
$\overline\cS$ from the interiors of calibrated rays, and that exclusion
fails for local flat models with all the support properties above. In
$1+1$ Minkowski space put
\[
 P=\{0\}\cup\{p_k=2^{-k}:k\ge1\},\qquad u_p=(\sqrt{1+p^2},p),
\]
and define the concave, globally Lipschitz function
\[
 b(t,z)=\min_{p\in P}\{\sqrt{1+p^2}\,t-pz+p^4\}.
\]
The minimum is attained by compactness of $P$. At every $(t,0)$, $t\ge0$,
the unique active parameter is $p=0$, so $b$ is differentiable there and
$\alpha(t)=(t,0)$ is a future calibrated ray.

Fix $t>0$ and set $F_t(p)=t\sqrt{1+p^2}+p^4$, a strictly convex function.
At
\[
 z_k(t)=\frac{F_t(p_k)-F_t(p_{k+1})}{p_k-p_{k+1}}
\]
the adjacent parameters $p_k,p_{k+1}$ are exactly the minimizers of
$F_t(p)-z_k(t)p$ over $P$, since the secant line through these two points
lies strictly below the graph at every other parameter of $P$, all of which
lie outside the open interval between them. Thus $(t,z_k(t))$ is singular,
while $z_k(t)\to F_t'(0)=0$. Every interior point of $\alpha$ belongs to
$\overline{\cS}\setminus\cS$.

Every active slope $u_p$ generates an immediately regular future calibrated
ray: for $s>0$ and $r\in P$, the affine function $L_r$ of index $r$
satisfies $L_r(q+su_p)=L_r(q)+s[-g(u_r,u_p)]\ge b(q)+s$, with equality for
the active branch $p$ and strict inequality for $r\ne p$ by the reverse
Cauchy--Schwarz inequality. This is a local eikonal model; it is not
asserted to arise from a compact Cauchy hypersurface. It shows that
closedness of $\cS$ near calibrated rays cannot be obtained from these
local properties alone, and is consistent with Corollary~\ref{cor:flat}.
\end{remark}

\section{An exact spacelike intrinsic arc}\label{app:example}

The sphere steps and the cone-causal curves above should be distinguished
from intrinsic Lax--Oleinik maximizing arcs. We give an explicit example,
with exact formulas, in which such an arc contains a spacelike segment.

On Minkowski space with coordinates $(t,x,y)$ and metric
$g=-\dd t^2+\dd x^2+\dd y^2$, set
\[
 u_a=(7/3,-2,2/3),\qquad u_m=(1,0,0),\qquad u_b=(7/3,2,2/3).
\]
These are future unit timelike vectors. Let $b_i(z)=-g(u_i,z)$ and, for
$\delta>0$, define
\[
 b(z)=\min\{b_a(z),b_m(z),b_b(z)+\delta\}.
\]
The function is globally Lipschitz and concave, with past unit gradient
where differentiable. At a point where branch $i$ is active, its future
ray with velocity $u_i$ is calibrated and is immediately regular. Indeed,
along that ray the active branch increases at unit rate, while every
distinct branch increases at rate $-g(u_j,u_i)>1$.
The triple stratum is the spacelike line
\begin{equation}\label{eq:triple}
 c=\{(\tau,\delta/4,2\tau+3\delta/4):\tau\in\R\}.
\end{equation}
The origin is singular, with branches $a,m$ active.

\subsection{The Hamiltonian and exact maximizing points}
Write $P=(P_0,P_1,P_2)$ and
\[
 K(P)=\exp\bigl(\sqrt{1+P_1^2+P_2^2}-P_0\bigr)-1.
\]
This Hamiltonian has positive definite fiber Hessian. To make it Tonelli,
add a smooth convex radial term $\rho(|P|^2-R^2)$, zero near
$\Delta=\co\{p_a,p_m,p_b\}$, where
\[
 p_a=(7/3,2,-2/3),\quad p_m=(1,0,0),\quad p_b=(7/3,-2,-2/3).
\]
Choose $\rho$ convex and nondecreasing, zero for nonpositive arguments,
and affine with positive slope for large arguments. The modified
$\widehat K$ is strictly convex and superlinear, and equals $K$ near
$\Delta$. The function $b$ is a viscosity solution of
$\widehat K(Db)=0$: the equation holds on the three branch differentials,
convexity gives the subsolution condition on their active convex hulls,
and at singular points the subdifferential is empty.

Let $L$ be the Legendre transform of $\widehat K$. Since the Hamiltonian is
independent of the base point, $A_s(0,z)=sL(z/s)$. The intrinsic arc is
\begin{equation}\label{eq:intrinsic}
 z_s=\argmax_z\{b(z)-sL(z/s)\}.
\end{equation}
It is unique because the maximized function is strictly concave and tends
to $-\infty$ at infinity. If $P\in D^+b(z)$ and
$z=s\widehat K_P(P)$, the supporting inequalities for $b$ and $L$ show
directly that $z$ is this unique maximizer.

For $1/2\le k\le2/3$, put
\[
 D(k)=45+24k-40k^2,
 \qquad
 P(k)=\left(1+\frac{12k}{\sqrt D},
                  \frac{6-2k}{\sqrt D},-\frac{6k}{\sqrt D}\right).
\]
One checks exactly that
\begin{align*}
 P(k)&=\alpha_a p_a+\alpha_m p_m+\alpha_b p_b,\\
 \alpha_a&=\frac{8k+3}{2\sqrt D},\qquad
 \alpha_m=1-\frac{9k}{\sqrt D},\qquad
 \alpha_b=\frac{10k-3}{2\sqrt D}.
\end{align*}
All three weights are positive and sum to one. Positivity of
$\alpha_m$ follows from
$D-81k^2=45+24k-121k^2\ge65/9>0$ on this interval.
Also
\[
 \sqrt{1+P_1(k)^2+P_2(k)^2}=\frac9{\sqrt{D(k)}}.
\]
Define
\begin{align}
 s(k)&=\frac{9\delta}{8(3-k)}
       \exp\left(1-\frac{9-12k}{\sqrt{D(k)}}\right),\label{eq:sk}\\
 z(k)&=\delta\left(-\frac9{8(3-k)},\frac14,
                         -\frac{3k}{4(3-k)}\right).\label{eq:zk}
\end{align}
Direct substitution gives $z(k)=s(k)K_P(P(k))$ and $z(k)\in c$.
Since all three branches are active there, $P(k)\in D^+b(z(k))=\Delta$.
Thus \eqref{eq:zk} is exactly the intrinsic arc at parameter
\eqref{eq:sk}. Moreover,
\[
 \frac{\dd}{\dd k}\log s(k)
 =\frac1{3-k}+\frac{216(3-k)}{D(k)^{3/2}}>0.
\]
The arc therefore runs through a nontrivial spacelike segment of $c$.
In particular its two points at $k=1/2$ and $k=2/3$ are
\[
 z(1/2)=\delta(-9/20,1/4,-3/20),\qquad
 z(2/3)=\delta(-27/56,1/4,-3/14),
\]
and their difference is $\delta(-9/280,0,-9/140)$, with squared Lorentzian
norm $243\delta^2/78400>0$. No reparameterization of this arc is causal.

For completeness, the same formulas extend over
$3/10<k<(12+3\sqrt{621})/121$, giving the full interior-triple-stratum
phase. The shorter rational interval used above already proves the claim
and requires no numerical certification.

\subsection{The past-slice maximization variant}
The variant maximizing $b(z)+d(z,0)$ on a fixed past time slice also has
a spacelike interval. Let $z=(\tau,\delta/4,2\tau+3\delta/4)\in c$ and
\[
 r=\sqrt{\tau^2-\delta^2/16-(2\tau+3\delta/4)^2}.
\]
The future unit direction from $z$ to the origin has spatial part
\[
 w_{\rm sp}=\left(-\frac{\delta}{4r},
                         \frac{-2\tau-3\delta/4}{r}\right).
\]
This belongs to the interior of the triangle
$\co\{(-2,2/3),(0,0),(2,2/3)\}$ precisely when
\begin{equation}\label{eq:sliceinterval}
 \frac{-39-\sqrt{69}}{96}\,\delta<\tau<-\frac5{12}\,\delta.
\end{equation}
Indeed the triangle inequalities are
$|w_1|<3w_2$ and $w_2<2/3$. The first gives $\tau<-5\delta/12$;
the second, after squaring within this range, gives
$192(\tau/\delta)^2+156(\tau/\delta)+121/4<0$, yielding the lower endpoint
in \eqref{eq:sliceinterval}. These inequalities also ensure $r>0$.
The spatial superdifferential of $b$ is the negative of that triangle,
while the spatial gradient of $d(\cdot,0)$ is $w_{\rm sp}$.
The optimality condition therefore holds at $z$. Concavity of $b$ and
strict concavity of the distance on the timelike part of the fixed slice
make $z$ the unique maximizer. As $\tau$ varies in
\eqref{eq:sliceinterval}, these maximizers trace a spacelike segment.
This is also the global maximum on the entire slice: outside the
timelike past of the origin, $d(\cdot,0)=0$ and $b\le b_m=\tau$, whereas
the displayed candidate has value $\tau+r>\tau$.

\subsection{What this example does and does not refute}
For $P$ near $\Delta$,
\[
 K_P(P)=e^{\sqrt{1+|P_{\rm sp}|^2}-P_0}
     \left(-1,\frac{P_{\rm sp}}{\sqrt{1+|P_{\rm sp}|^2}}\right)
\]
is past timelike. Thus every generalized characteristic satisfying
$\dot z\in\co\widehat K_P(D^+b(z))$ is past timelike on compact ranges.
The intrinsic arc just constructed cannot satisfy that inclusion, even
after reparameterization.

The data here are globally Lipschitz, globally concave and solve a
base-independent Tonelli equation. Their Lipschitz and semiconcavity
bounds are independent of $\delta$. Scaling $\delta$ puts the spacelike
phase within any prescribed uniform short-time interval. Consequently
this example contradicts the identification of the intrinsic maximizing
arc with the generalized characteristic in
\cite[Theorem~3.6, equation~(3.19)]{CC}. It does not by itself refute a
separate existential statement about other singular characteristics.
It is also a local eikonal model, not a construction of a nonsmooth global
Cauchy Busemann function on a spatially compact spacetime.

\begingroup\raggedright

\endgroup
\end{document}